\documentclass[sigconf]{acmart}
\usepackage{algorithm}
\usepackage{algorithmic}

\usepackage{xspace}
\newcommand*{\eg}{\emph{e.g.\@\xspace}}
\newcommand*{\ie}{\emph{i.e.\@\xspace}}

\usepackage{amsmath}
\DeclareMathOperator*{\maximize}{\text{\fontfamily{pcr}\selectfont maximize}}

\usepackage{mathtools}
\usepackage{amsfonts}
\usepackage{subfig}
\usepackage{mathrsfs}
\usepackage{dsfont}
\usepackage{multirow}

\AtBeginDocument{%
  \providecommand\BibTeX{{%
    \normalfont B\kern-0.5em{\scshape i\kern-0.25em b}\kern-0.8em\TeX}}}

\copyrightyear{2021} 
\acmYear{2021} 
\setcopyright{acmlicensed}\acmConference[KDD '21]{Proceedings of the 27th ACM SIGKDD Conference on Knowledge Discovery and Data Mining}{August 14--18, 2021}{Virtual Event, Singapore}
\acmBooktitle{Proceedings of the 27th ACM SIGKDD Conference on Knowledge Discovery and Data Mining (KDD '21), August 14--18, 2021, Virtual Event, Singapore}
\acmPrice{15.00}
\acmDOI{10.1145/3447548.3467103}
\acmISBN{978-1-4503-8332-5/21/08}

\begin{document}
\fancyhead{}

\title{Neural Auction: End-to-End Learning of Auction Mechanisms for E-Commerce Advertising}
\author{Xiangyu Liu$^{1*}$, Chuan Yu$^{1*}$, Zhilin Zhang$^1$, Zhenzhe Zheng$^{2\dagger}$, Yu Rong$^1$, Hongtao Lv$^2$, Da Huo$^2$, Yiqing Wang$^2$, Dagui Chen$^1$, Jian Xu$^1$, Fan Wu$^2$, Guihai Chen$^2$ and Xiaoqiang Zhu$^1$}
\affiliation{
  \institution{$^{1}$Alibaba Group, China, $^{2}$Shanghai Jiao Tong University, China
  \and $^{1}$\{qilin.lxy,yuchuan.yc,zhangzhilin.pt,homer.ry,dagui.cdg,xiyu.xj,xiaoqiang.zxq\}@alibaba-inc.com
  \and $^{2}$\{zhengzhenzhe,lvhongtao,sjtuhuoda,wangyiqing\_2015\}@sjtu.edu.cn, \{fwu,gchen\}@cs.sjtu.edu.cn}
  \country{}
}

% \renewcommand{\shortauthors}{Xiangyu Liu, Chuan Yu, \emph{et al}.}

\begin{abstract}

%% 玺羽版本：
%In e-commerce advertising, it is crucial to jointly consider various performance metrics, \eg, user experience, advertiser utility, and platform revenue. Traditional auction mechanisms, such as GSP and VCG auctions, can be suboptimal due to their fixed allocation rules to optimize a single performance metric (\eg, revenue or social welfare). Recently, data-driven auctions, learned directly from auction outcomes to optimize multiple performance metrics, have attracted increasing research interests. However, the procedure of auction mechanisms involves various discrete calculation operations, making it challenging to be compatible with continuous optimization pipelines in machine learning. In this paper, we design \underline{D}eep \underline{N}eural \underline{A}uctions (DNAs) to enable end-to-end auction learning. More specifically, deep models are developed to efficiently extract and compress data from auctions, providing rich features for auction design. A differentiable model is proposed to relax the sorting operation, a key component in auctions, to a continuous operation. This is the key to enabling end-to-end learning of auctions. The conditions for game theoretical properties have also been explicitly expressed in the model design. DNAs have been successfully deployed in the e-commerce advertising system at Taobao. Experimental evaluation results on both large-scale data set as well as online A/B test demonstrated that DNAs significantly outperformed other widely adopted mechanisms in the industry.
%%

% % 郑老师版本：
In e-commerce advertising, it is crucial to jointly consider various performance metrics, \eg, user experience, advertiser utility, and platform revenue. Traditional auction mechanisms, such as GSP and VCG auctions, can be suboptimal due to their fixed allocation rules to optimize a single performance metric (\eg, revenue or social welfare). Recently, data-driven auctions, learned directly from auction outcomes to optimize multiple performance metrics, have attracted increasing research interests. However, the procedure of auction mechanisms involves various discrete calculation operations, making it challenging to be compatible with continuous optimization pipelines in machine learning. In this paper, we design \underline{D}eep \underline{N}eural \underline{A}uctions (DNAs) to enable end-to-end auction learning by proposing a differentiable model to relax the discrete sorting operation, a key component in auctions. 
We optimize the performance metrics by developing deep models to efficiently extract contexts from auctions, providing rich features for auction design.
We further integrate the game theoretical conditions within the model design, to guarantee the stability of the auctions. DNAs have been successfully deployed in the e-commerce advertising system at Taobao. Experimental evaluation results on both large-scale data set as well as online A/B test demonstrated that DNAs significantly outperformed other mechanisms widely adopted in industry.
% %

{\let\thefootnote\relax\footnote{
{$^{*}$Equal contribution. $^{\dagger}$Corresponding author.}}}
\end{abstract}



\maketitle

\section{Introduction}
\label{sec:intro}

In online e-commerce, the advertising platform is an intermediary to help advertisers deliver their products to interested users~\cite{goldfarb2011online}. 
Auction mechanisms, such as
 Vickrey-Clarke–Groves (VCG) auction~\cite{vickrey1961counterspeculation}, Myerson auction~\cite{myerson1981optimal} and generalized second-price auction (GSP)~\cite{edelman2007internet},
have been used to enable  efficient ad allocation in various advertising scenarios. 
%matching between advertisers and users~\cite{lucking2000vickrey}. 
On designing auction mechanisms for e-commerce advertising, we need to jointly consider and optimize multiple performance metrics from three major stakeholders, \emph{i.e.}, users, advertisers, and the ad platform. Users look for good shopping experiences, advertisers want to accomplish their ad marketing objectives, and the ad platform would like to extract large revenue while also provide satisfying services to both users and advertisers~\cite{bachrach2014optimising,zhang2021optimizing}. Furthermore, the ad platform may balance and adjust the importance of these metrics to satisfy the business's strategies for users and advertisers in different e-commerce scenarios.
%of e-commerce advertising. 
%, the importance of these metrics may be adjusted according to the business's strategy. 
High-quality user experiences and advertising services would guarantee the long-term prosperity of e-commerce advertising. 
The traditional auction mechanisms are suboptimal for the e-commerce advertising with multiple performance metrics in dynamic environments. VCG auction and Myerson auction focus on optimizing either social welfare or revenue, and the procedures of the auctions are also too complex to explain for advertisers. Although GSP auction has nice interpretation and is easy to deploy in industry, 
%and achieves  success in industry, 
the fixed allocation rule limits its capability to optimize multiple performance metrics in dynamic environments. 
%has  interpretation. 
%The VCG auction suffers from high computational complexity and, while the Myerson auction relies on the exact knowledge about value distributions, which are seldom available in practice. 
%, such as VCG auction~\cite{vickrey1961counterspeculation}, Myerson auction~\cite{myerson1981optimal} and GSP auction~\cite{edelman2007internet} are not longer optimal for the e-commerce advertising with multiple optimization objectives whose importance weights are adjustable.    
%The VCG auction suffers from high computational complexity, while the Myerson auction relies on the exact knowledge about value distributions, which are seldom available in practice. 
To overcome these limitations, we turn our attention to data-driven auction mechanisms, inspired by the recent increase of interest on leveraging modern machine learning, and in particular deep learning, for auction design~\cite{duetting2019optimal,feng2018deep,rahme2021auction,shen2019automated}.
The data-driven auction mechanisms enable us to exploit rich information, such as the context of auction environment and the performance feedback from auction outcomes, to guide the design of flexible allocation rules for optimizing multiple performance metrics, 
%The data-driven approaches
which significantly enlarge the design space of auction mechanisms.

\begin{figure}[!t]
\centering
\includegraphics[width=0.47\textwidth]{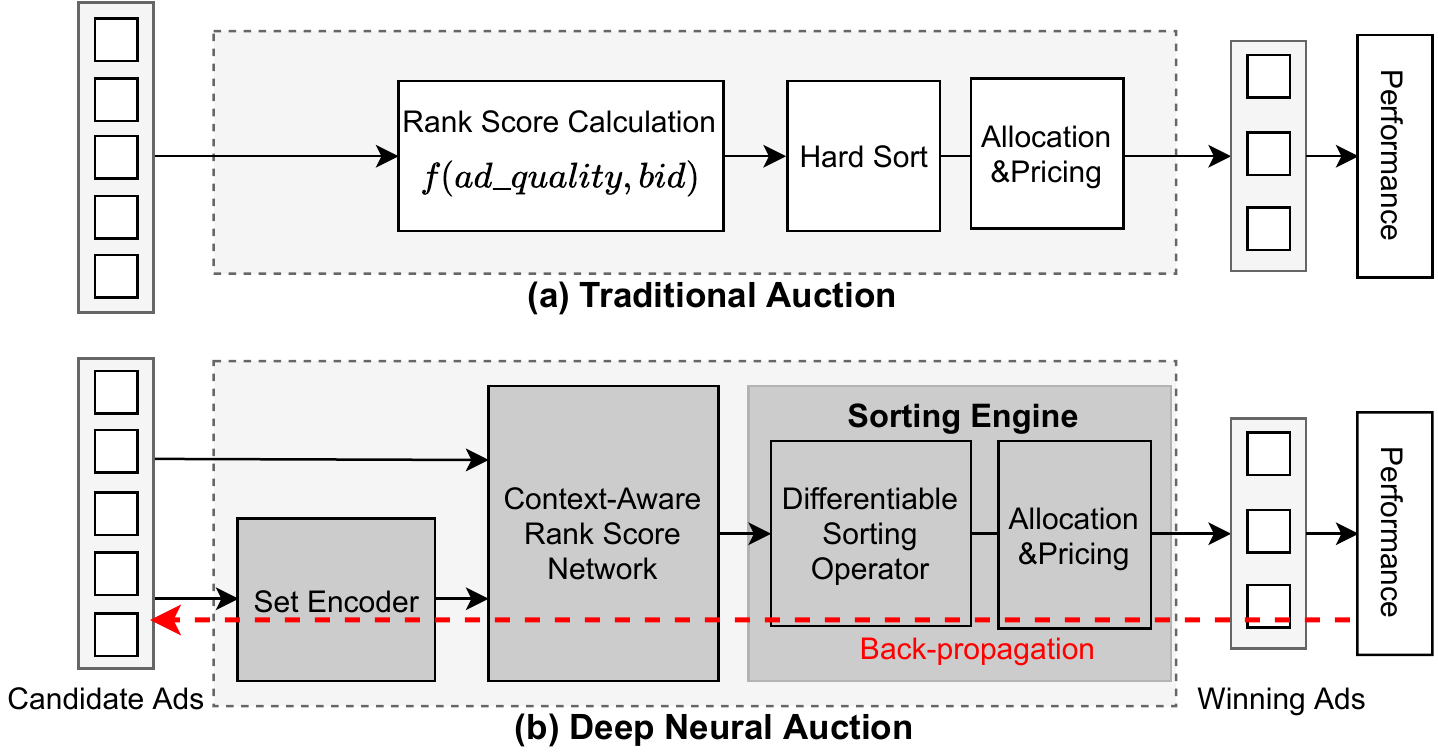}
\caption{The comparison between traditional auctions and Deep Neural Auction.
The set encoder and the  context-aware rank score network are applied to extract auction features, which improves representation space and flexibility of rank scores, compared with the fixed rank score in traditional auctions.
Furthermore, the differentiable sorting engine makes the auctions, including allocation and pricing, continuous and differentiable w.r.t the inputs, thereby supporting the end-to-end back-propagation training.
}
\label{fig:dna_architecture}
\end{figure}

However, it remains open to both academia and industry on how to make full use of the powerful deep learning on designing data-driven auction mechanisms for the industrial e-commerce advertising. 
We consider two critical challenges in this direction. 
The first one comes from the contradiction between auction mechanism and deep learning in design principle.
%We stick to
The auction mechanisms, including allocation and pricing, usually involve various discrete optimization operations, 
%rely on a sorting operation to derive integer-valued solutions, 
\eg, the top-k ads selection in GSP auction~\cite{varian2007position}, 
%(the procedures of allocation and payment) usually rely on a sorting operation, 
while the deep learning follows an end-to-end pipeline for continuous optimization. 
This contradiction prevents the performance feedback underlying the auction outcomes from integrating into the back-propagation model training as in deep learning. 
%Accompanying is 
When designing learning models for data-driven auctions, we also need to take the game theoretical properties, such as \emph{Incentive Compatibility}~\cite{vickrey1961counterspeculation}, into account, which further complicates the application of deep learning for auction design.  
%to the auction design. 
%is a poor match for the end-to-end differentiable pipelines in deep learning.
%, as the allocation procedure and the calculation of corresponding payments are not natively differentiable. 
%Specially, both  allocation and payment in auctions rely on a
%the key component of an auction, 
%sorting operation,
%to calculate an integer-valued solution, 
%which is actually not differentiable and prohibits the gradient-based optimization from deep learning. 
%greedy allocation (i.e., sorting by rank scores) 
%is which prohibits end-to-end gradient-based optimization. 
%The auction mechanism cannot be integrated as a differentiable component in deep neural networks, which will further exacerbate the sample-efficiency issue.
Although some recent works have proposed deep neural network architectures for learning-based auction mechanisms~\cite{duetting2019optimal, feng2018deep, shen2019automated}, they focused on the theoretical auction setting, either the complex combinatorial auctions~\cite{duetting2019optimal} or the simple single-bidder auctions~\cite{shen2019automated}, in lack of the insights from industrial deployment. Thus, it needs further efforts to integrate deep learning into the design and deployment of end-to-end auction mechanisms for practical industrial setting. %especially for the multi-slot setting. 
%for practical e-commerce advertising scenarios, such as . 
%adapt them to the real applications in e-commerce advertising, especially for the widely used multi-slot auctions. 

%. for this direction. 
%The existing learning-based auction mechanisms typically follow an \emph{individual virtual value}\footnote{This name follows the concept of virtual value from the optimal auction design~\cite{myerson1981optimal}.} transform: which 
%virtual value prediction} scheme: designing a new transformation, which 
%maps the ad related features together with the bid to a new rank score, from which the allocation and payment can be derived. 
%The transform extends the simple rank score in the GSP. 
%The extending the simple rank score in the GSP. 
%from which the allocation and payment can be derived. 
%and the truthful property is maintained. 
%This framework, however,  faces two critical challenges in applying deep learning for auction design. 
%The current learning-based mechanisms~\cite{} may not be very 
The second challenge is data efficiency. The current learning-based approaches~\cite{zhang2021optimizing,golrezaei2017boosted} usually require a large number of samples to learn the optimal auction due to an ambiguity issue we observed in the data from auctions.
%consider a common case in e-commerce advertising,
It is a common case that an advertiser with the same feature profile can result in different outcomes in distinct auctions, \emph{e.g.}, wins in one auction but loses in another, due to the change of the auction context, \emph{e.g.}, the competition from the other advertisers. %be a bidder in two distinct auctions, but
%she may get entirely different outcomes (e.g., she wins the auction in the first auction but loses in the second). 
From the machine learning perspective, this may cause an ambiguity issue~\cite{hullermeier2006learning}, introducing the one-to-many relation in samples, \emph{i.e.}, the same feature (advertiser feature profile) but contradictory labels (winning or losing). 
%if the individual features of her ad remain unchanged, as there may contains one-to-many relationship samples (i.e. same features but contradictory labels) in the training data set.
% as the machine learning model will always predict the same labels with which it has been trained.
%This inefficiency is the characteristic of 
% As neural network models do not incorporate much inductive bias~\cite{battaglia2018relational}, the ambiguity phenomenon leads to the inefficiency in data samples.
Naive neural network models, which do not incorporate much inductive bias~\cite{battaglia2018relational}, may not fully handle the ambiguity phenomenon on data samples from auctions, resulting in inefficient learning for the end-to-end auction design.

%the problem of sample .  
%which may cause the aforementioned sample-efficiency issue. 

In this paper, we aim to develop a data-efficient end-to-end \underline{D}eep \underline{N}eural \underline{A}uction mechanism, namely DNA, to optimize multiple performance metrics for e-commerce advertising. 
%make an in-depth study towards learning data efficient and end-to-end auction mechanisms
%for the industrial e-commerce advertising. 
Considering the nice properties of easy interpretation and deployment in industry, we stick to the rank score-based allocation and second-price payment procedures inherited from GSP auction. 
%for the multiple ad slots setting, and 
%the rank score-based allocation and second-price payment procedures from GSP auction. 
%We consider the general multiple ad slots setting, and
%We focus our discussions on the widely used GSP auction for the multi-slot setting with the objective of optimizing multiple performance metrics~\cite{bachrach2014optimising,zhang2020optimizing}.  
%widely used second-price multi-slot auctions when optimizing multiple performance metrics, which is an essential research topic in e-commerce advertising~\cite{bachrach2014optimising,zhang2020optimizing}.
 %which takes the input of a set of advertisers with rich features, calculates a context-aware rank score under the guideline of performance metrics, and outputs the winning advertisers
%permutation over the ad set, i.e. allocation,
%as well as the corresponding second-price payments. 
%This deep neural auction mechanism
%learning-based rank score design ratio functi
%d the design rationale of DNA 
As shown in Figure~\ref{fig:dna_architecture}, DNA contains a set encoder, a family of context-aware rank score functions and a differentiable sorting engine, with which the process of auction design can be integrated into an end-to-end learning pipeline. 
Specifically, the set encoder and a carefully designed neural network extract and compress the rich features of auction, such as auction context, bid, advertiser's profile, predicted auction outcomes, etc, into a context-aware rank score, tremendously increasing the representation space and flexibility of rank scores in GSP auction. 
%As the sorting operation is not differentiable, 
We further propose a module to relax the sorting in auctions as a differentiable operation, which makes the whole process of GSP auction, including allocation and pricing, to be continuous and fully differentiable with respect to the inputs, and then supports the end-to-end  back-propagation training. 
When designing the learning models for these three modules, we introduce several constraints over the network structures, such as the monotonicity of the neural network in terms of bid, to preserve the game theoretical properties of GSP auction for DNA. Our contributions in this paper can be summarized as follows:

$\bullet$ We make an in-depth study on leveraging the power of deep learning to design data-driven 
    %data efficient and end-to-end
    auctions for industrial e-commerce advertising. The proposed end-to-end Deep Neural Auction mechanisms, namely DNA, enable us to optimize multiple performance metrics using the real feedback from historical auction outcomes. The newly designed rank scores also largely enhance the flexibility of ad allocation, making it possible to adjust the auction mechanisms for various performance metrics in dynamic environments. 
    
    %in industrial e-commerce advertising, 
    %consider the input candidate ads as a whole via set embedding learning and models the interrelationship between them in ranking, which makes them more competitive than other learning-based ad auction mechanisms based on only ad rank score functions.
    
    $\bullet$ We employ three deep learning models to facilitate the design of data efficient and end-to-end learning auction mechanisms with the guarantee of game theoretical property. %We consider the candidate ads as a whole, and 
    %A set encoder model,
    A set encoder model and 
    a monotone neural network model are proposed to encode various features of auction into the context-aware rank score. 
    %extract the context of auction environment, which is attached to the features of advertisers, to resolve the ambiguity issue of data samples.  
    %from to dthe features of all candidate ads, the . 
    %We further propose an explicit rank score model t
    %and models the interrelationship between tem in ranking, 
    With the proposed differentiable sorting engine, we can formulate the design of data-driven auction as a continuous optimization problem, which can be integrated into an end-to-end learning pipeline. 
    %owe can integrate the whole process of GSP auction (allocation and payment)
    %(concretely, sorting and payment operations) 
    %fully compatible with differentiability,
 %into an end-to-end neural network architecture by constructing a differential sort operation, 
    %making the discrete sorting procedure 
  %  which enables gradient back-propagation in an end-to-end leaning pipeline.
    
    $\bullet$ We have deployed the DNA mechanism in the advertising system at \emph{Taobao}, one of the world's leading e-commerce advertising platforms. Experimental results on both large-scale industrial data set as well as the online A/B test showed that DNA mechanism significantly outperformed other widely used industrial auction mechanisms on optimizing multiple performance metrics, such as Utility-based GSP~\cite{bachrach2014optimising} and Deep GSP~\cite{zhang2021optimizing}. 
% \end{itemize}

%  Secondly, it constrains the network architecture using partially monotone MIN-MAX network to guarantee the truthful property for e-commerce value maximizers, which is simple but does not restrict the learning capacity. Last but not the least, with the help of continuous relaxations, Deep Neural Auction is enabled to model the whole process of greedy allocation inside the neural network by making the discrete sorting procedure compatible with differentiability, which facilitates back-propagation and end-to-end learning.

% Experimental results on large-scale industrial e-commerce data set as well as the online A/B test in an advertising system showed that Deep Neural Auction significantly outperformed the baselines on optimizing multiple performance metrics, including the traditional auction mechanisms, such as GSP, utility-based GSP, and a recent proposed learning-based auction model, Deep GSP~\cite{zhang2020optimizing}. 
% Empirical analyses also showed that the mechanism learned by Deep Neural Auction is robust to the input orders and sizes, making it to be a stable ranking model. \todo{The last sentence will be removed.}

% The rest of this paper is organized as follows. Section~
% \vspace{-1.0em}
\section{Preliminaries}
\label{sec:preliminaries}
\subsection{Ad Auction Model}
We describe a typical ad platform architecture in e-commerce advertising.
Formally, $N$ advertisers compete for $K \leq N$ ad slots, which are incurred by a PV (page view) request from the user. Each advertiser $i$ submits bid $b_i$ based on her private information, which could be the probability of the user's behaviors (\eg, $pCTR$, etc.) over the ad, obtained by learning-based prediction module~\cite{cheng2016wide,zhou2018deep}. 
We use vector $\mathbf{b}=(b_i,\mathbf{b}_{-i})$ to represent the bids of all advertisers, where $\mathbf{b}_{-i}$ are the bids from all advertisers except $i$.
 We represent the ad auction mechanism by $\mathcal{M}\langle\mathcal{R},\mathcal{P}\rangle$, where $\mathcal{R}$ is the ad allocation scheme and $\mathcal{P}$ is the payment rule. 
 The ad allocation scheme would jointly consider the bids and the quality (\eg, $pCTR$ and $pCVR$) of the ads in a principled manner. We use $\mathcal{R}_i(b_i, \mathbf{b}_{-i})=k$ to denote the advertiser $i$ wins the $k$th ad slot, while $\mathcal{R}_i(b_i, \mathbf{b}_{-i})=0$ represents the advertiser loses the auction. 
 The $K$ winning ads would be displayed to the user. 
The auction mechanism module further calculates the payments for the winning ads with a rule $\mathcal{P}$, which would be carefully designed to guarantee the economic properties and the revenue of the auction mechanism.  

\subsection{Problem Formulation}
% As introduced in Section~\ref{sec:intro}, the performance metrics in e-commerce advertising can be diver, dynamic and possibly conflicting with each other. 
% In this paper, we focus on designing an auction mechanism to optimize multiple performance metrics in e-commerce advertising. Users, advertisers and ad platform all have their own objectives. Users look for good shopping experiences, advertisers want to accomplish their marketing objectives, and the ad platform would like to extract high revenue in providing satisfying services to both users and advertisers. For the long-term prosperity, one critical tool the ad platform can use to jointly optimize the above mentioned multiple objectives (i.e., performance metrics) is the auction mechanism. The auction mechanism determines the ads displayed to users as well as the payments charged to advertisers.

Follow the work~\cite{zhang2021optimizing}, we formulate the problem as \emph{multiple performance metrics optimization in the competitive advertising environments}. Given bid vector $\textbf{b}$ from all the advertisers and $L$ ad performance metric functions ${\{f_1(\mathbf{b};\mathcal{M}), .. , f_L(\mathbf{b};\mathcal{M})\}}$ (such as Revenue, CTR, CVR, etc), we aim to design an auction mechanism $\mathcal{M}\langle \mathcal{R}, \mathcal{P} \rangle$, such that
% ${\{f_j(\mathbf{b};\mathcal{M})\}}_{1}^{L}$,
\begin{equation}
% \small
\begin{aligned}
\maximize_{\mathcal{M}} \quad & \mathbb{E}_{\mathbf{b} \sim \mathcal{D}} [F(\mathbf{b};\mathcal{M})]\\
\textrm{s.t.} \quad 
& \textit{Incentive Compatibility (IC) constraint,}\\
& \textit{Individual Rationality (IR) constraint,}\\
\end{aligned}
\label{eq:problem}
\end{equation}
where $\mathcal{D}$ is the advertisers' bid distribution based on which bidding vectors $\mathbf{b}$ are drawn. We define $F(\mathbf{b};\mathcal{M}) = \lambda_1 \times f_1(\mathbf{b};\mathcal{M}) + \cdots + \lambda_L\times f_L(\mathbf{b};\mathcal{M})$, where the objective is to maximize a linear combination of the multiple performance metrics $f_l$'s with preference parameters $\lambda_l$'s. 
% We can design different auction mechanisms to make various trade-offs among performance metrics by choosing different preference parameters. There are a series of related works on how to determine $\lambda_j$ and derive Pareto-efficient solutions~\cite{lin2019pareto,xiao2017fairness,chen2017optimizing}. These discussions are beyond the scope of this work, and
The parameters $\lambda_l$'s are the inputs of our problem. The constraints of IC and IR guarantee that advertisers would  truthfully report the bid, and would not be charged more than their maximum willing-to-pay for the allocation, which are important for the stability of the ad auction and would be discussed in details in Section~\ref{sec:ic}.

In this work, we stick to the design rationale of classical GSP auction mechanism~\cite{lahaie2007revenue,bachrach2014optimising}, where the allocation scheme is to rank advertisers according to their \emph{rank scores} with a non-increasing order. The pricing rule is to charge the winning advertisers with the minimum bid required to maintain the same ad slot.
We study a learning-based GSP auction framework, leveraging the power of deep neural network to design a new rank score function and integrate it into the GSP.
% Specifically, we build a deep neural network to map advertiser's bid to a rank score, with the consideration of various related information, such as ad features (ads category, $pCTR$, and $pCVR$), user profile (gender, age, and income) and advertiser preference (budget, marketing demands). 
We use $r(b_i, \mathbf{x}_i)$ to denote this new rank score function, where $(b_i, \mathbf{x}_i)$ denotes all available information, including bid and other features related to the advertiser $i$'s ad, in the auction. For ease of presentation, we also denote it as $r_i(b_i)$ if there is no ambiguity.
The training of this non-linear model is under the guideline of optimization objective in~(\ref{eq:problem}). With this new rank score, the allocation scheme and payment rule can be summarized as follows:
%\footnote{we slightly abuse the subscript $i$ as both the $i$th advertiser and the advertiser at the $i$th slot, if there is no ambiguity.} 
% \begin{itemize}

    $\bullet$ Allocation Scheme $\mathcal{R}$: Advertisers are sorted in a non-increasing order of new rank score $r_i(b_{i})$. Without loss of generality, let
    \begin{equation}\label{eq:allocation}
    % \small
    r_1(b_{1})\geq r_2(b_{2}) \geq \cdots \geq r_N(b_{N}),
    \end{equation}
    %\begin{equation}\label{eq:rank_score}
    %r_{i}=R_\theta(b_{i}),\mbox{ where } \nabla_{b}R_{\theta}(b_i) \geq 0.
    % r_{i}=R_\theta(b_{i}),\mbox{ where } \frac{\partial R_{\theta}(b_{i})}{\partial b_{i}} \geq 0.
    %\end{equation}
    then the advertisers with top-K scores would win the corresponding ad slots, with ties broken randomly.
    
    $\bullet$ Payment Rule $\mathcal{P}$: The payment for the winning advertiser $i$ is calculated by the formula:
    \begin{equation}\label{eq:pricing}
    % \small
    % P: p_{j}=R_{\theta}^{-1}(r_{j}^{-}) + \Delta,
    p_{i}=r_i^{-1}(r_{i+1}(b_{i+1})),
    \end{equation}
    where $r_{i+1}(b_{i+1})$ is the rank score of the next highest advertiser, and $r_{i}^{-1}(\cdot)$ is the inversion function of $r_{i}(\cdot)$.
% \end{itemize}

% The objective is to maximize a linear combination of the multiple performance metrics $\{f_j\}^L_1$ with preference parameters. By choosing different $w_j$'s, we can design auction mechanisms to make various trade-offs among performance metrics. In this paper, we assume $w_j$'s are the inputs in the problem formulation, and focus on the ad auction mechanism design. There are extensive related works on how to determine $w_j$'s and derive Pareto-efficient solutions~\cite{lin2019pareto,chen2017optimizing}.

% \subsection{Learning-based Ad Auctions}

% \subsection{Optimal Mechanism Design with Interdependent Values}
% \label{sec:interdependent}
% \todo{xxx}

% In practice, interdependent values are a more realistic model of bidders’ values than independent values\todo{cite}. Interdependence captures correlated private values, which arise when bidders’ information about their value for winning the auction is correlated with that of others~\cite{roughgarden2016optimal}.

% We use interdependence to model that advertisers do not know their precise value for winning an auction since it depends on others’ information. The model of interdependence thus enriches the set of underlying assumptions that we are able to make about the informational structure of the auction setting. In this paper, we explicitly treat such informational assumptions and their roles in designing learning-based ad auction mechanisms, by incorporating a separate deep neural network module to extract this informational structure on candidate ad set.

\subsection{Economic Properties}
\label{sec:ic}
In the auction mechanism design, one cannot just assume that an advertiser $i$ would truthfully reveal her maximum willing-to-pay price $m_i$ in the auction\footnote{$m_i$ is not necessarily equal to the value $v_i$ of the PV request. For example, there may be budget constraints.}, since they have incentives to misreport these prices to manipulate their own interests~\cite{edelman2007strategic}. This may seriously harm the stability of the advertising platform. Therefore, we need to guarantee the property of \emph{incentive compatibility} (IC), from mechanism design. This property removes the computational burden of bidding strategy optimization from advertisers, and also, leads to reliable and predictable inputs for the auction mechanisms.
\begin{definition}[Incentive Compatibility~\cite{vickrey1961counterspeculation}]
An auction mechanism is IC if it is in the best interest of each advertiser $i$ to truthfully reveal her maximum willing-to-pay price, \emph{i.e.}, $b_i=m_i$.
\end{definition}

% \footnote{To motivate why $v_i$ and $m_i$ might be different, consider buying a house with a higher value than the amount of money your bank is willing to lend you.}
%on ad performance optimization.

%Given the advertisers' bids, an auction can decide how much to charge the different advertisers per engagement. Unfortunately, one cannot just assume that the advertisers will reveal their true value in the auction, if this information is eventually used to determine the payments. Indeed, in many auction settings, participants have an incentive to under-report their values to reduce the expected payment derived by the auction~\cite{edelman2007strategic}. However, when the auction mechanism is truthful, it incentivizes its participants to reveal their true value to the auction mechanism. It would also remove the burden of considering advertisers’ strategic behaviors, leading to reliable and predictable inputs for ad performance optimization. This gives us a strong motivation to design truthful auctions in our settings.

In traditional auction theory, the celebrated IC auction mechanisms, such as VCG auction~\cite{vickrey1961counterspeculation} and Myerson auction~\cite{myerson1981optimal}, typically build upon the assumption that advertisers are \emph{utility maximizers}, that is, the goal of each advertiser $i$ is to optimize her quasi-linear utility, defined as the difference between her expected value $v_i$ and the payment $p_i$, \ie, $u_i = v_i - p_i$. However, we observe from the industrial e-commerce platform that this model could not fully capture the behavior pattern of advertisers. 
For example, in \emph{Taobao} advertising platform, 
%In an industrial e-commerce advertising platform, 
there are two representative types of advertisers:
%One with upper bounds of bids, and another with constraints over the average costs, such as pay-per-click (PPC), pay-per-acquisition (PPA). 
% \todo{add some references}
Optimized Cost Per Click (OCPC) advertisers with upper bounds of bids, and Multi-variable Constrained Bidding (MCB) advertisers with constraints over budgets and the average costs, such as pay-per-click (PPC) and pay-per-acquisition (PPA). 
The goal of both types of advertisers is to optimize the overall realized value of advertising, such as the quantity of conversions and clicks, under certain constraints over the payments. For these types of advertisers, they would calculate and report a maximum willing-to-pay price for each PV request based on the current status of the ad campaign, with the help of auto-bidding services~\cite{zhu2017optimized,yang2019bid}. 
This behavior pattern of advertisers could be well captured by the model of \emph{value maximizer}~\cite{wilkens2017gsp}, which is defined as follows:
\begin{definition}[Value maximizer~\cite{wilkens2017gsp}]
A value maximizer $i$ optimizes value $v_i$ while keeping payment $p_i$ below her maximum willing-to-pay $m_i$; when value is equal, a lower $p_i$ is preferred.
\end{definition}

In the auctions with multiple ad slots, a value maximizer prefers to the outcome with a higher slot when the payment is below the maximum willing-to-pay price, and then a smaller payment is preferred under the situation with equal value. 
%We hence could obtain that 
The strategic behavior pattern of value maximizers would be quite different from the traditional utility maximizers. 
%For example, they may seek for lower indexed slots with disproportionally larger costs.
%However, in the industrial e-commerce advertising, the behaviors of advertisers are complicated. Given an impression opportunity, their true value can hardly be calculated precisely, which is different from that in the celebrated auction theory~\todo{cite}. Furthermore, besides the widely-used utility from game theory, advertisers typically have diverse optimization goals, such as GMV or ROI. To optimize these diverse objectives, each advertiser submits a basic (average) bid and a budget for multiple auction instances, and an intelligent agent helps advertiser to adjust the actual bid for each auction instance based on the context of advertising environments~\cite{zhu2017optimized,wu2018budget}. After checking the advertising performance of auctions during a certain period, the advertisers adjust their bids, budgets and other decision variables accordingly. We find that this phenomenon is well captured by the concept of ``value maximizer'' proposed by~\citeauthor{wilkens2017gsp}~\cite{wilkens2017gsp}. A value maximizer has a preference over outcomes and a maximum willingness to pay for each outcome. She strategizes to achieve the most preferred outcome without incurring a price above what she is willing to pay.
It has been proved that an auction mechanism is IC for value maximizers, as long as the following two conditions are satisfied~\cite{aggarwal2009general, wilkens2017gsp}: 

$\bullet$ \textbf{Monotonicity}: An advertiser would win the same or a higher slot if she reports a higher bid; 

$\bullet$ \textbf{Critical price}: The payment for the winning advertiser is the minimum bid that she needs to report to maintain the same slot. 

We note that IR is also guaranteed under these two IC conditions. Obviously given the monotonicity constraint, the critical price is strictly lower than the bid, and hence is lower than the maximum willing-to-pay price, \ie, $p_i < m_i$. %Thus the auction does not charge a bidder more than her expected value.
It could be easily verified that GSP satisfies these conditions and hence is IC and IR for value maximizers. 
In this work, we would design learning-based auction mechanisms, following the above conditions.

\section{Deep Neural Auction}
In this section, we present the details of Deep Neural Auction (DNA) mechanism for optimizing multiple performance metrics under the multi-slot setting for e-commerce advertising.

\begin{figure*}[!t]
\captionsetup[subfigure]{justification=centering}
\centering
\subfloat[Deep Neural Auction Architecture]{
    \label{fig:pipeline}
	\includegraphics[width=0.75\textwidth]{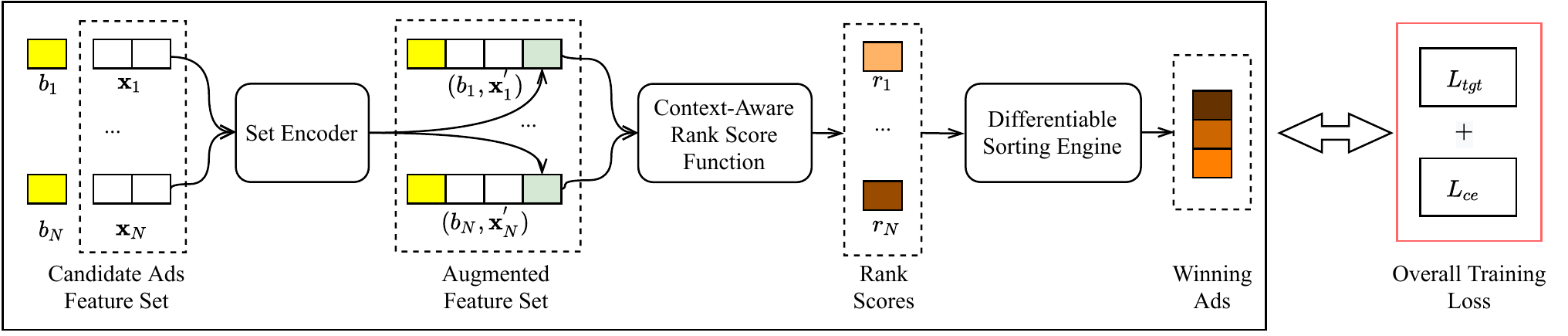}}
\\
\subfloat[Set Encoder: Deep Set Architecture][Set Encoder:\\ Deep Set Architecture]{
	\label{fig:deepset}
	\includegraphics[width=0.32\textwidth]{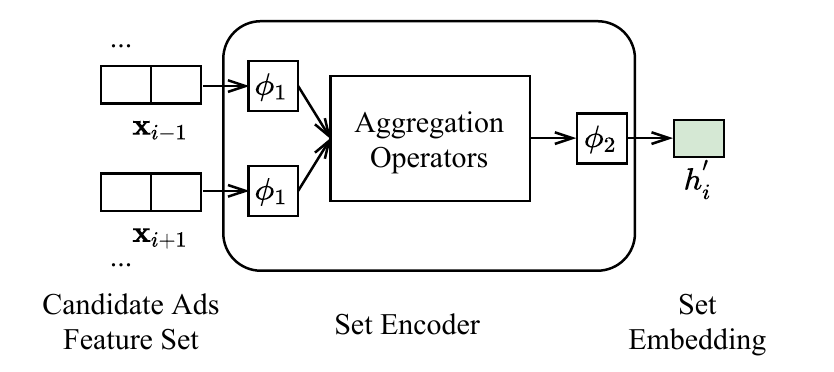}}
\subfloat[Context-Aware Rank Score Function: Partially Monotone MIN-MAX Network][Context-Aware Rank Score Function:\\ Partially Monotone MIN-MAX Network]{
	\label{fig:minmax}
	\includegraphics[width=0.32\textwidth]{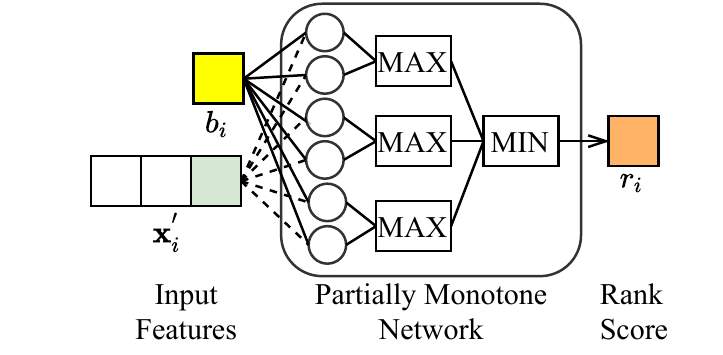}}
\subfloat[Differentiable Sorting Engine: NeuralSort Module][Differentiable Sorting Engine:\\ NeuralSort Module]{
	\label{fig:engine}
	\includegraphics[width=0.32\textwidth]{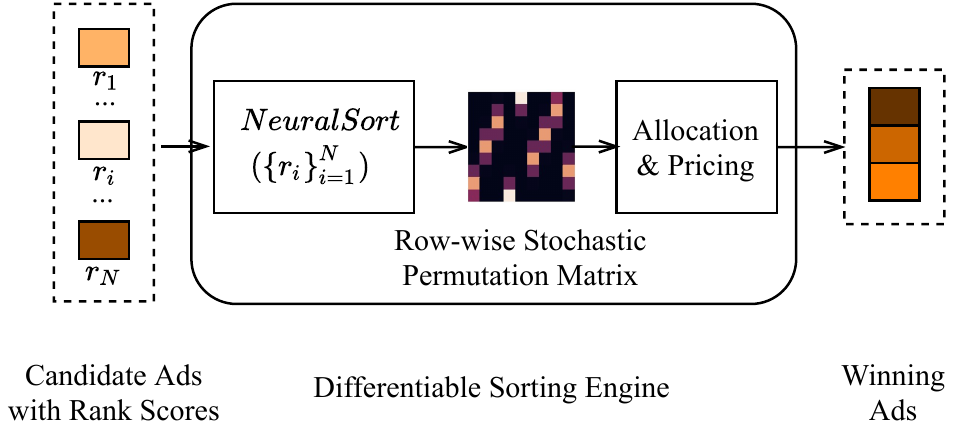}
}
\caption{(a) The overall Deep Neural Auction architecture. (b) The Deep Set based set encoder receives the whole set of ad features and outputs a set embedding. (c) Partially monotone MIN-MAX neural network based context-aware rank score function. The straight lines represent connections with non-negative weights, whereas the dashed lines represent unconstrained connections. (d) The differentiable sorting engine takes in the generated rank scores and outputs a row-stochastic permutation matrix of {\fontfamily{pcr}\selectfont argsort} as well as its corresponding allocation and payments, by using NeuralSort.}
\label{fig:all_model}
\end{figure*}

% \begin{figure*}[!t]
% \centering
% \includegraphics[width=0.33\linewidth]{pic/neuralsort.pdf}
% % \vspace{-0.5em}
% \caption{}
% % \vspace{-0.5em}
% \label{fig:model}
% \end{figure*}

% \begin{figure*}[!t]
% \centering
% \includegraphics[scale=0.49]{pic/model.pdf}
% % \vspace{-0.5em}
% \caption{(a) The architecture of deep set, which receives the whole set of ad features and outputs a permutation-invariant embedding. (b) The architecture of a partially monotone MIN-MAX neural network. The straight lines represent connections with non-negative weights, whereas the dashed lines represent unconstrained connections. (c) The overall Deep Neural Auction structure. Best viewed in colour.}
% % \vspace{-0.5em}
% \label{fig:model}
% \end{figure*}

\subsection{Overall Architecture}
As illustrated in Fig.~\ref{fig:pipeline}, DNA consists of three modules: a set encoder, a context-aware rank score function, and a differentiable sorting engine. The set encoder learns a set embedding from the features of candidate ads, which encodes the context of the auction. This set embedding is attached as a complementary feature for each ad.
% The set encoder uses DeepSet architecture to learn a permutation invariant set embedding on can ad set.
%a partial monotonic MIN-MAX neural network.
Next, each advertiser employs a shared MIN-MAX neural network to generate context-aware rank scores from advertiser's features. 
This neural network is partially monotonic with respective to bids, which is critical to the guarantee of IC property. 
Another advantage of the designed neural network is a closed form expression for the inverse transform, which enables an easy payment calculation. 
%which also supports an closed-form expression for second-price payment calculation. 
Then, the differentiable sorting engine conducts a continuous relaxation of sorting operator in auctions, 
%over all the generated rank scores, 
and outputs a row-stochastic permutation matrix. We can use this row-stochastic permutation matrix to express the expected revenue as well as other predicted performance metrics, from which training losses from real feedback underlying the auction outcome can be constructed. 
With these components, the whole DNA mechanism is fully differentiable with respect to its inputs, and can be integrated into the end-to-end learning pipeline as in deep learning.  
%and is enabled to
%learn data efficient and end-to-end auction mechanisms for optimizing multiple performance metrics. 
% second-price auction procedure, by using continuous relaxation of discontinuous sorting operator, and making the sorting process differentiable with respect to its inputs. 
% This differentiable allocation engine outputs a row-stochastic permutation matrix, which will be used to derive a task-specific training loss from real feedback. The whole DNA framework is enabled to learn the optimal auction for optimizing multiple performance metrics in an end-to-end learning-based procedure.
It is worth to note that both the set encoder and the context-aware rank score function are parameterized neural network models, while the differentiable sorting engine is a non-parameterized operator. We next introduce the details of these three modules. 
%The following sections will introduce the details of these three modules, as well as the model training.

% We first describe the motivation of modeling all bidders in candidate ad set due to the inspiration from interdependent value settings in mechanism design~\cite{roughgarden2016optimal}, and leverage deep set~\cite{zaheer2017deep} to auto encode the session information.

\vspace{-0.8em}
\subsection{Auction Context Encoding}
As the final auction outcome is jointly determined by all the candidate ads, we design a set encoder to automatically extract the feature of the auction context from all the candidate ads, instead of the individual ad. 
We attach this auction context feature as an augmented feature for each ad to overcome the ambiguity issue discussed in Section~\ref{sec:intro}.

% As discussed in Section~\ref{sec:intro}, an \emph{individual virtual value} transformation model may suffer from the ambiguity issue with the learning-based auction design approaches.
% A natural solution is to explicitly extract the context of auctions from the candidate ads to help it disambiguate between different auction outcomes.
% As the final auction outcome is determined by all candidate advertisers, we design a set encoder to automatically extract features from the set of candidate ads features in this auction.

% However, this feature set has some inherent informational structure. Intuitively, the advertisers are indistinguishable and permutation-equivariant, although individual advertisers are entirely different (\eg \textit{\{bidder-1, bidder-2, bidder-3\}} and \textit{\{bidder-3, bidder-2, bidder-1\}} are totally equivalent for the allocation in ad auction). Such auctions are anonymous and invariant to relabeling the advertisers, i.e. they can be executed without any information about the advertisers, or labeling them. This holds in e-commerce advertising, as there is no relative ordering implied between any two distinct eligible ads.

The set encoder receives the whole set of ad features as input. As there is no inherent ordering among the advertisers in the set, the feature set is permutation-equivariant~\cite{zaheer2017deep} to the auction outcome, \ie, the auction outcome does not rely on the ordering of the features. 
% However, this poses a challenge for machine learning models.
Learning models that do not take this set structure into account (such as MLPs or RNNs) would cause the issue of discontinuities ~\cite{zhang2019fspool}.
% We define the \textit{advertisers-symmetric} in e-commerce advertising:
% \begin{customdef}{1}[\protect{Advertisers-Symmetric}]\label{def:advertiser_symmetric}
% For each PV request from the user, the N advertisers' joint valuation distribution $D$ is advertisers-symmetric if for any permutation of the advertisers $\varphi_b$, the permuted distribution $D_{\varphi_b} \coloneqq D_{\varphi_b(1)} \times \cdots \times D_{\varphi_b(N)}$ satisfies: $D_{\varphi_b} = D$.
% \end{customdef}
% Intuitively, the definition of advertiser-symmetric means that the advertisers are indistinguishable and permutation-equivariant, although individual advertisers are entirely different (\eg \textit{\{bidder-1, bidder-2, bidder-3\}} and \textit{\{bidder-3, bidder-2, bidder-1\}} are totally equivalent for the allocation in ad auction). Such auctions are anonymous and invariant to relabeling the advertisers, i.e. they can be executed without any information about the advertisers, or labeling them. This holds in e-commerce advertising, as there is no relative ordering implied between any two distinct eligible ads. However, this poses a challenge on modeling set using machine learning models, since there is no inherent ordering to the elements in the set. Models that do not take this set structure into account properly (such as MLPs or RNNs) will result in discontinuities issue\todo{some cites}.
Inspired from the recent progress on learning on set~\cite{zhang2019fspool, zaheer2017deep}, we implement the set encoder by designing a new deep neural network, which uses DeepSet~\cite{zaheer2017deep} network to aggregate individual ad features to form a representation for the auction context. 
%This set encoder generates a representation from the set of candidate ads by aggregating individual ad representations, and uses the DeepSet~\cite{zaheer2017deep} architecture for this aggregation. 
The main idea is that, by setting equivariant layers and a final symmetric layer, the DeepSet can learn to aggregate all the features' information in a permutation-equivariant manner. The generated embedding vector can be trained to predict some interested statistical value about the whole set. 
% such as the mean or median of all the bids. 

% Concretely, as illustrated in Fig.~\ref{fig:deepset}, the set encoder is composed of two layers. Given the features set of candidate ads $\{(b_i, \mathbf{x}_i)\}_{i=1}^{N}$, each instance $(b_i, \mathbf{x}_i)$ is firstly mapped to a high-dimensional latent space through a shared fully connected layer $\phi_1$, resulting in a set of intermediate hidden states $\mathbf{h} = \{h_i\}^{N}_{i=1}$:
Concretely, as illustrated in Fig.~\ref{fig:deepset}, the set encoder is composed of two groups of layers, $\phi_1$ and $\phi_2$. Given the features of candidate ads $\{\mathbf{x}_i\}_{i=1}^{N}$, each instance $\mathbf{x}_i$ is firstly mapped to a high-dimensional latent space through shared fully connected layers $\phi_1$, resulting in a set of intermediate hidden states $\mathbf{h} = \{h_i\}^{N}_{i=1}$:
\begin{equation}\label{eq:deepset1}
% \small
    h_i = \sigma(\phi_1(\mathbf{x}_i)),
\end{equation}
where $\sigma$ is an Exponential Linear Unit (ELU) activation function~\cite{DBLP:journals/corr/ClevertUH15}. Then, this hidden states set is processed with symmetric aggregation pooling (such as average pooling) to build the final set embedding $h'_i$ for each ad $i$ with another fully connected layer $\phi_2$:
\begin{equation}\label{eq:deepset2}
% \small
h'_i=\sigma(\phi_2(\text{\fontfamily{pcr}\selectfont avgpool}(\mathbf{h}_{-i}))),
\end{equation}
where $\mathbf{h}_{-i}$ represent the hidden states from all advertisers except $i$. 
%To summarize, 
This set encoder is built by composing permutation-equivariant operations (shared nonlinear transformation) with symmetric aggregation operations (average pooling) at the end. 
% It is easy to verify the permutation-invariant property because changing the order of elements in the input set will not affect the output feature embedding.
Since the symmetric operations are commutative to the input items, the output is the same regardless of the order of the items. The set encoder learns to extract the context of auction on the whole set of candidate ads~\cite{zaheer2017deep,DBLP:conf/iclr/EdwardsS17}, which is driven by the downstream training signals in an end-to-end manner. The output set embedding would be sent to the downstream rank score function as an augmented feature for each ad, which helps to infer each ad's rank score in the current candidate ads set.
It should be noted that the set encoder does not include the bids from all candidate ads, as shown in Fig.~\ref{fig:deepset}. This design is specified mainly for the guarantee of IC property, keeping the affection of each ad's rank score only through her bid, which will be elaborated in Section~\ref{sec:minmax}. 

% We also emphasize that the set encoder architecture does not require hard-coding the number of ads during training. Somewhat surprisingly, we will show in the Experiment part that training Deep Neural Auction model on instances with $50$ ads generalizes well even to $400$ ads in real-life e-commerce ad auction.

\subsection{Context-Aware Rank Score}
\label{sec:minmax}
%The context-aware rank score function takes 
We design a deep neural network to transform each advertiser's  augmented feature to a context-aware rank score. 
We use $r(b_i, \mathbf{x}'_i)$ to denote this rank score, where $\mathbf{x}'_i$ represents the augmented features except bid, \ie, $\mathbf{x}'_i = (\mathbf{x}_i, h'_i)$.
% For ease of presentation, we also denote it as $r_i(b_i)$ if there is no ambiguity.
From Section~\ref{sec:ic}, we need to satisfy two conditions to guarantee the IC property for value maximizers: monotone allocation and critical price.
Thus, we aim to design a strictly monotone neural network with respect to bid, and supports efficient inverse transform given the next highest rank score.
%which also supports calculating its corresponding inverse given the next highest rank score.

% this rank score function should satisfy the monotone allocation and critical bid based pricing. Thus designing a truthful auction for value maximizers reduces to finding a strictly monotone transformation on bid, which also supports calculating its corresponding inverse given the next highest rank score.

We incorporate the aforementioned constraints within the network architecture design, and restrict the search space to a family of partially monotone parameterized neural network. We model the rank score function as a two-layer feed-forward network with {\fontfamily{pcr}\selectfont min} and {\fontfamily{pcr}\selectfont max} operations over linear functions~\cite{daniels2010monotone} as shown Fig.~\ref{fig:minmax}. For $Q$ groups of $Z$ linear functions, we associate strictly positive weights $e^{w_{qz}}$ with $b_i$ and other unconstrained weights $w'_{qz}$ with $\mathbf{x}'_i$, as well as intercepts $\alpha_{qz}$, where $q=1,...,Q, z=1,...,Z$. For simplicity, we denote $r(b_i, \mathbf{x}'_i)$ as $r_i$, and assume $r_1\ge r_2\ge ... \ge r_N$
without loss of generality. We can define:
% \footnote{$w'_{qz}$ is in vector form.}
\begin{equation}
\label{eq:minmax_forward}
% \small
r_i = \min_{q\in [Q]}\max_{z\in [Z]}(e^{w_{qz}} \times b_i + w'_{qz} \times \mathbf{x}'_i + \alpha_{qz}).
\end{equation}
Since each of the above linear function in Eq.~(\ref{eq:minmax_forward}) is strictly non-decreasing on $b_i$, so is $r_i$. 
This partially monotone MIN-MAX neural network has been proved with the capability to approximate any  function~\cite{daniels2010monotone}. Another particular advantage of this representation is that the inverse transform can be directly obtained from the parameters for the forward transform in a closed form expression. For example, given the next highest rank score $r_{i+1}$, the payment for advertiser $i$ can be formulated as follows:
\begin{equation}
\label{eq:minmax_inv}
% \small
p_i = \max_{z\in[Z]}\min_{q\in[Q]}e^{-w_{qz}}(r_{i+1}-\alpha_{qz}-w'_{qz} \times \mathbf{x}'_i).
\end{equation}

With the above designed MIN-MAX neural network, the two conditions for IC property can be satisfied, given the assumption that the bids affect the rank scores only through the input $b_i$ in $r(b_i, {\bf x}'_i)$. However in the industrial advertising environment, there are some engineered features in $\mathbf{x}_i$ from domain knowledge that may have complex dependence relation with the bids. Therefore the bids may affect the rank scores and then the allocation in a complex way, which may violate this assumption.
%For example, the bidding information encoded in the context feature may affect the calculation of the rank score and the inverse transform of the payment. 
In a large-scale advertising platform, this effect of a change of one's bid on the rank scores via this route would be quite small from our observations on the industrial data sets. 
%Although the advertisers may have certain chance to benefit from bid manipulation in this scenario, they have to figure out the complex dependence relation of bids and rank scores in the black-box learning model, and thus is hard to deploy in practice. 
To investigate the influence of this issue on IC property, we conduct comprehensive experiments in Section~\ref{sec:exp_ic} to calculate the data-driven IC metric of DNA for value maximizers. We reserve the discussion about the strictly IC DNA mechanism design as an interesting open problem in our future work.

\subsection{Differentiable Sorting Engine}

% With the rank score function defined in Section~\ref{sec:minmax}, the allocation scheme and payment rule for multi-slot auctions can be summarized as follows:
% \begin{itemize}
%     \item Allocation Scheme $\mathcal{R}$: Advertisers are sorted in a non-increasing order of new rank score $r_i(b_{i})$:
%     \begin{equation}
%     r_1(b_{1})\geq r_2(b_{2}) \geq \cdots \geq r_N(b_{N}).
%     \end{equation}
%     The advertisers with the top-$K$ scores would win this auction.
%     \item Pricing Rule $\mathcal{P}$: The payment for each winning advertiser $i$ is calculated by Eq.~(\ref{eq:minmax_inv}).
% \end{itemize}

% These two procedures, however are agnostic to the proposed model, as they are deployed on an online advertising engine. 

After calculating the rank scores of all ads, the mechanism determines the allocation and payment, following Eq.~(\ref{eq:allocation}) and~(\ref{eq:pricing}).
% which are done at the online advertising engine in industrial ad platforms.
However, treating allocation and payment outside the model learning (\ie, as an agnostic environment) is in some sense poorly suited for deep learning. That is the processes of allocation and payment (actually the sorting operation) are not natively differentiable, and the gradients must all be evaluated via finite difference or likelihood ratio methods (such as the policy search used in Deep GSP~\cite{zhang2021optimizing}), with some additional issues of convergence stability and data-efficiency. Also, in another line of related work, the model-based reinforcement learning (RL) has achieved some notable successes~\cite{moerland2020model}. Some recent works used a general neural network to learn a differentiable dynamic model~\cite{kurutach2018modelensemble} and argued that the model-based approaches are often more superior and data-efficient than model-free RL methods for many tasks~\cite{ha2018world,de2018end}. These insights give us a motivation to model the whole process of allocation and payment inside the Deep Neural Auction framework.

In various types of auction mechanisms, both the allocation and payment are built on a basic sorting operation.
% However, sorting is a poor match for the end-to-end, automatically differentiable pipelines in deep learning.
Sorting operation outputs two vectors, neither of which is differentiable. 
On the one hand, the vector of sorted values is piecewise linear.
On the other hand, the sorting permutation (more specifically, the vector of ranks via {\fontfamily{pcr}\selectfont argsort} operator) also has no differentiable properties as it is integer-valued.

To overcome this issue, we propose a differentiable sorting engine that caters to the top-$K$ selection in the multi-slot auctions. We present a novel use of differentiable sorting operator, \ie, NeuralSort~\cite{DBLP:conf/iclr/GroverWZE19}, to derive a differentiable top-$K$ permutation matrix, which can be used to generate the various expected outcomes of the auctions. Given a set of unsorted rank scores
%\footnote{$\mathbf{r}$ is represented as a vector form just for ease of presentation. There is no inherent order between these rank scores.}
$\mathbf{r}=[r_1,r_2,\cdots,r_N]^T$, we are concerned with the {\fontfamily{pcr}\selectfont argsort} operator, where {\fontfamily{pcr}\selectfont argsort}$(\mathbf{r})$ returns the permutation that sorts $\mathbf{r}$ in a decreasing order.
%\footnote{This is an adaption to our top-$K$ selection in multi-slot auctions, which is different from the conventional definition of {\fontfamily{pcr}\selectfont argsort} (which is in an increasing order).}. 
Formally, we define the {\fontfamily{pcr}\selectfont argsort} operator as the mapping from $N$-dimensional real vectors $\mathbf{r}\in\mathbb{R}^{N}$ to the permutations over $N$ elements, where the %{\fontfamily{pcr}\selectfont argsort} 
permutation matrix $M_{r} \in \{0,1\}^{N\times N}$ is expressed as  
\begin{equation}
M_{r}[k,i] =
\begin{cases}
1 & \text{if $i=$ {\fontfamily{pcr}\selectfont argsort}$(\mathbf{r})[k]$,}\\
0 & \text{otherwise.}
\end{cases}
\end{equation}
Here $M_{r}[k,i]$ indicates if $r_i$ is the $k$th largest rank score in $\mathbf{r}$. The results from ~\cite{DBLP:conf/iclr/GroverWZE19} showed the identity:
\begin{equation}
\label{eq:sort}
    M_{r}[k,i] = 
    \begin{cases}
    1 & \text{if $i=$ {\fontfamily{pcr}\selectfont argmax}$(c_k)$,}\\
    0 & \text{otherwise,}
    \end{cases}
\end{equation}
where $c_k=(N+1-2k)\mathbf{r}-A_{r}\mathds{1}$, $A_{r}$ denotes the absolute pairwise differences of elements in $\mathbf{r}$ such that $A_{r}[i,j]=|r_i - r_j|$, and $\mathds{1}$ denotes the column vector of all ones. Then, by relaxing the operator {\fontfamily{pcr}\selectfont argmax} in Eq.~(\ref{eq:sort}) by a row-wise {\fontfamily{pcr}\selectfont softmax}, we can arrive at the following
continuous relaxation for the {\fontfamily{pcr}\selectfont argsort} operator $M_r$, which is called NeuralSort in~\cite{DBLP:conf/iclr/GroverWZE19}:
\begin{equation}
\small
\label{eq:sort2}
    \hat{M}_{r}[k,:] = \text{\fontfamily{pcr}\selectfont softmax}(\frac{c_k}{\tau}),
\end{equation}
where $\tau > 0$ is a temperature parameter that controls the degree of the approximation, and as $\tau \rightarrow 0$, $\hat{M}_{r} \rightarrow M_{r}$. 
Intuitively, the $k$th row of $\hat{M}_r$ can be interpreted as the `choice' probabilities on all elements in $\mathbf{r}$, for getting the $k$th highest item.

This row-stochastic permutation matrix $\hat{M}_{r}$, can be used as a basic operator to construct task-specific sorting procedures according to the order of generated rank scores in a differentiable manner. For instance, if we let $\mathbf{p}=[p_1,p_2,\cdots,p_N]^T$ denotes the payments calculated by Eq.~(\ref{eq:minmax_inv}) for $N$ advertisers in a PV request, then the top-$K$ payments, sorted by their corresponding rank scores, can be recovered by a simple matrix multiplication:
\begin{equation}
\label{eq:F_pay}
\small
    f_{pay} = \hat{M}_r{[1{:}K,:]} \cdot \mathbf{p}.
\end{equation}
This row-stochastic permutation matrix $\hat{M}_r$ acts as a differentiable sorting engine that makes the discrete sorting procedure compatible with differentiability.
% and connects the parametrized models with the real feedback, which will be introduced in the next section.

% As the online ad platform feeds all the inputs from an ad set (i.e. all ads' features) to the neural auction model for multi-slots auction at a time, our deep neural auction model makes allocation decision for a single .

\subsection{End-to-End Model Training}
% The input of Deep Auction Model is a set of candidate ads for an ad request, and the output is the row-stochastic permutation over the set.

% \textbf{Data for Training}:
\subsubsection{Data for Training}
% We emphasize that the data used for training must be under a truthful auction, only then learning ad auctions from data can be meaningful. Therefore,
All data sets we used were generated under GSP auction, which is IC for e-commerce value maximizing advertisers~\cite{wilkens2017gsp}.
The data contains all advertisers' bids, the estimated values (\eg, $pCTR$, $pCVR$), ads information (\eg, category, price of product), user features (\eg, genders, age, income level) as well as the context information (\eg, the source of traffic). These information consists of the input features of the DNA architecture. The data also contains the real feedback information (\eg, click, conversion or transaction) from users.
% The features of each ad would reflect the quality of ad opportunity and the status of auction environments. We consider the following information to represent: 1) Ad information, such as bid, $pCTR$, $pCVR$, and ad category. 2) Advertisers' information, like the current budget, the price of products, and marketing intent. 3) User features, such as gender, age, income level, shopping preferences, etc.

% \textbf{Training Loss}:
\subsubsection{Training Loss}
As the training data contains the user feedback for each ad exposure, we can directly use the row-stochastic permutation matrix $\hat{M}_r$ to compute the $K$-slots expected performance metrics via: $\hat{M}_r[1{:}K,:] \cdot F_{all}$, where $F_{all}$ represents the vector of aggregated performance metrics for all ads from real feedback:
\begin{equation}
\label{eq:f_all}
\small
F_{all}=[\sum_{l=1}^{L}\lambda_l\times f^1_l,\cdots, \sum_{l=1}^{L}\lambda_l\times f^N_l]^T,
\end{equation}
with $f^i_l$ standing for the $l$th performance metric for $i$th ad from a PV request. Therefore, we can formulate the learning problem as minimizing the sum of top-$K$ expected negated performance metrics for each PV request:
\begin{equation}
\label{eq:loss1}
    \mathcal{L}_{tgt} = -\sum_{i=1}^{K} \hat{M}_r[i,:] \cdot F_{all}.
\end{equation}
One exception is the calculation of revenue. Due to the change of allocation order, the payment for each ad is distinct from what has happened. Thus we use the generated payments, defined in Eq.~(\ref{eq:F_pay}) to replace the ones appeared in the training data.

We set another auxiliary task to help train the DNA mechanism. With the benefit of hindsight from real feedback, we can access the optimal allocation to maximize the performance metrics in each PV request. Thus we set another multiclass prediction task, whose loss is the row-wise cross-entropy (CE) between the ground-truth and the predicted row-stochastic $N\times N$ permutation matrix:
\begin{equation}
\label{eq:loss2}
    \mathcal{L}_{ce} = -\frac{1}{N}\sum^{N}_{k=1}\sum^{N}_{i=1}\mathds{1}(M_y[k,i]=1)\text{\fontfamily{pcr}\selectfont log} \hat{M}_r[k,i],
\end{equation}
where $M_y$ is the ground-truth permutation matrix, calculated by sorting their real feedback. We found that this auxiliary task was beneficial to yield a stable training process in our experiments. We use a hyper-parameter to balance the target loss $\mathcal{L}_{tgt}$ and the cross-entropy term $\mathcal{L}_{ce}$.
% However, the sparse issue is common in real industrial e-commerce advertising. Since users typically decide to purchase a product long after seeing an ad, the user feedback, especially with respect to the conversion behaviors, are scarce and delayed.

However, the user feedback, especially with respect to the conversion behaviors, is scarce in industrial e-commerce advertising, \eg, users typically decide to purchase a product after seeing dozens of ads. To alleviate this problem, we replace the sparse user behaviors (typically one-hot) in data with the dense values from the prediction model (such as $pCTR$ and $pCVR$), and debias these predicted values with real user behaviors by the \emph{calibration} techniques~\cite{borisov2018calibration,deng2021calibrating}. We also eliminate the deviation of CTR between different slots by debiasing with the posterior inherent CTR of different slots.

% To alleviate this problem, calibration \cite{deng2021calibrating,  su2020attention, borisov2018calibration}, as a simple but effective method, has been widely-adopted in the industry, but 
% to our best knowledge, it has attracted rare attention from the academic. On the other hand, the feedback from advertisers, such as the satisfaction and activity, are also sparse and delayed. For example, an advertiser may leave the ad platform if she is unsatisfied with the ad performance for a long time. Taobao has also conducted some  primary explorations on this problem \cite{guo2020deep}.

\section{Experimental Evaluations}
\label{sec:exp}
% In this section, we firstly introduce the experimental setup, including the evaluation metrics and baselines. Then we conduct offline simulations to evaluate the performance comparison with baseline mechanisms. Finally, we deploy the proposed DNA mechanism on a real e-commerce ad platform, and collect the evaluation results.

\subsection{Experiment Setup}

\subsubsection{Evaluation Metrics}
We consider the following metrics in our offline and online experiments, which reflect the platform revenue, user experience, as well as advertisers' utility in the e-commerce advertising. For all experiments in this paper, metrics are normalized to a same scale.
% For all the experiments in this paper, metrics are scaled to $[0,1]$, without loss of generality.

\textbf{1) Revenue Per Mille (RPM).}
$RPM = \frac{\sum click\times PPC}{\sum impression}\times1000$.

\textbf{2) Click-Through Rate (CTR).}
$CTR = \frac{\sum click}{\sum impression}$.

\textbf{3) Conversion Rate (CVR).}
$CVR = \frac{\sum order}{\sum impression}$.

\textbf{4) GMV Per Mille (GPM).}
$GPM = \frac{\sum merchandise\mbox{ }volume}{\sum impression} \times 1000$.

Apart from the advertising performance indicators, we also evaluate the effectiveness of our designed learning-based auction mechanisms on the property of IC.

\textbf{5) IC Metric ($\boldsymbol{\Psi}$).} We propose a new data-driven metric of IC, $\boldsymbol{\Psi}$, to represent the ex-post regret of value maximizers, similar to the data-driven IC for utility maximizers \cite{feng2019online}. %Based on the definition of value maximizer, we denote 
The metric of $\boldsymbol{\Psi}$ consists of the regret on value $\boldsymbol{\Psi}_v$ and the regret on payment $\boldsymbol{\Psi}_p$,
which denotes, through bid perturbation, the maximum percentage of  value increase under the payment constraint, and the maximum percentage of payment decrease with the identical allocation, respectively.
%under payment constraint 
%as a tuple consisting of the regret on value $\boldsymbol{\Psi}_v$ and the regret on payment $\boldsymbol{\Psi}_p$,  
Concretely, we formulate $\boldsymbol{\Psi}=(\boldsymbol{\Psi}_v, \boldsymbol{\Psi}_p)$ as\footnote{Since our training data comes from a vanilla GSP mechanism, which is IC for value maximizers, we directly take the bid $b_i$ as the maximum willing-to-pay price $m_i$ of advertiser $i$.}:
\begin{equation}
\label{psi_v}
    \boldsymbol{\Psi}_v = \frac{1}{N}\sum_{i=1}^{N}\frac{1}{\beta_{k_i}} \max_{b_i'}((\beta_{k'_i} - \beta_{k_i})\times \mathds{1}(p_i(b'_i)<m_i)),
\end{equation}
\begin{equation}
    \boldsymbol{\Psi}_p = \frac{1}{N}\sum_{i=1}^{N}\frac{1}{\beta_{k_i} p_i(b_i)}
    \max_{b_i'}((\beta_{k_i} p_i(b_i) - \beta_{k'_i} p_i(b'_i))\times \mathds{1}(k'_i=k_i)),
\end{equation}
where we denote $k_i$ and $k'_i$ as the allocated slot indexes of advertiser $i$ when bidding truthfully and bidding a perturbed $b'_i$, respectively. We use $\beta_k$ as the click-through rate of slot $k$, and $p_i(b_i)$ as the payment of advertiser $i$ when bidding $b_i$. 
It should be noted that the value of a click $v_i$ for advertiser $i$ is reduced from the fraction in Eq.~(\ref{psi_v}). 
%We recall the reader that $m_i$ denotes the maximum willing-to-pay price for advertiser $i$.
Intuitively, $\boldsymbol{\Psi}$ measures to what extent a value-maximizing advertiser could get better off via manipulating her bid. 
A larger value of $\boldsymbol{\Psi}_v$ indicates that an advertiser could obtain more extra value under the constraint of maximum willing-to-pay price. Similarly, a larger value of $\boldsymbol{\Psi}_p$ indicates that an advertiser could be undercharged more while obtaining the same ad slot.
For example, as GSP is IC for value maximizers, its values of both $\boldsymbol{\Psi}_v$ and $\boldsymbol{\Psi}_p$ are 0.

%As the industrial ad auction system does not support counterfactual evaluations of the auction's outcomes for alternative bids~\cite{deng2020data}, we leverage a data-driven metric, IC-Envy~\cite{colini2020envy}, to quantify equilibrium using intermediate data with only black-box access to the auction mechanism. Let $\bm{\beta}$ be the expected click-through rate vector of all ad slots and $\mathcal{P}(v_i, b_{-i})$ be the expected payment vector of all bidders, the IC-Envy metric is defined as:
% \begin{equation}
% \begin{split}
%     \mbox{IC-Envy}(v_i) = max_{j}&(\beta_j\times v_{i}-p_j(v_i,b_{-i}))\\
%     -&(\beta_{i}\times v_{i} - p_i(v_i, b_{-i})).
% \end{split}
% \end{equation}
% IC-Envy measures to what extend advertisers are happy with the outcome. An envy-free allocation is that all bidders prefer the slot they receive over any other slots (given prices). Further analysis of IC-Envy can be found in~\cite{colini2020envy}.

% \subsection{Offline Experiments}
\subsubsection{Baselines Methods} We compare DNA with the widely used mechanisms in the industrial ad platform.

\textbf{1) Generalized Second Price auction (GSP).} The rank score in the classical GSP is simply the bids multiplying $pCTR$, namely effective Cost Per Milles (eCPM). The payment rule is the value of the minimum bid required to retain the same slot. The work~\cite{lahaie2007revenue} suggested incorporating a squashing exponent $\sigma$ into the rank score function, \ie, $bid \times pCTR^{\sigma}$ could improve the performance, where $\sigma$ can be adjusted to weight the performance of revenue and CTR. We refer to this exponential form extension as GSP in the experiments.

\textbf{2) Utility-based Generalized Second Price auction (uGSP).} uGSP extends the conventional GSP by taking the rank score as a linear combination of multiple performance metrics using estimated values: $r_i(b_i) = \lambda_1\times b_i\times pCTR_i + o_i$, where $o_i$ represents other utilities, such as CTR and CVR: $o_i=\lambda_2\times pCTR_i + \lambda_3\times pCVR_i (\mbox{where }\lambda_l \geq 0)$. The payment of uGSP follows the principle from GSP: $p_i = \frac{\lambda_1\times b_{i+1}\times pCTR_{i+1} +o_{i+1} - o_{i}}{\lambda_1 \times pCTR_i}$. uGSP is  widely used in industry to optimize multiple performance metrics~\cite{bachrach2014optimising}.

\textbf{3) Deep GSP~\cite{zhang2021optimizing}} Deep GSP uses a deep neural network to map ad's related features to a new rank score within the GSP auction. This new rank score function is optimized using model-free reinforcement learning to maximize the interested performance metrics.

%\textbf{4) Generalized First Price auction (GFP).} GFP allocates the ad slots greedily based on eCPM, and the payment of a winning advertiser is simply her bid. Since GFP is obviously not an incentive compatible mechanism, we do not compare its performance with DNA, and only adopt it in the comparisons of incentive compatibility.

\subsection{Offline Experiments}
\subsubsection{Datasets}

The data sets we used for experiments come from \emph{Taobao}, a leading e-commerce advertising system in China. We randomly select 5 million records logged data under GSP auctions from \textit{July 4, 2020} as training data, and 870k records logged data from \textit{July 5, 2020} as test data. Unless stated otherwise, all experiments are conducted under the setting of top-3 ads displayed (\ie, 3-slot auctions) in each PV request. Other details about the model configurations and training procedure are in Appendix~\ref{app:training}.
% We built an offline auction simulator with a prediction module to generate simulated feedback and implemented the baseline mechanisms in this simulator. 

\subsubsection{Performance in Offline Simulations}
\label{sec:performance_offline}

\begin{figure}[!t]
\centering
\includegraphics[width=0.39\textwidth]{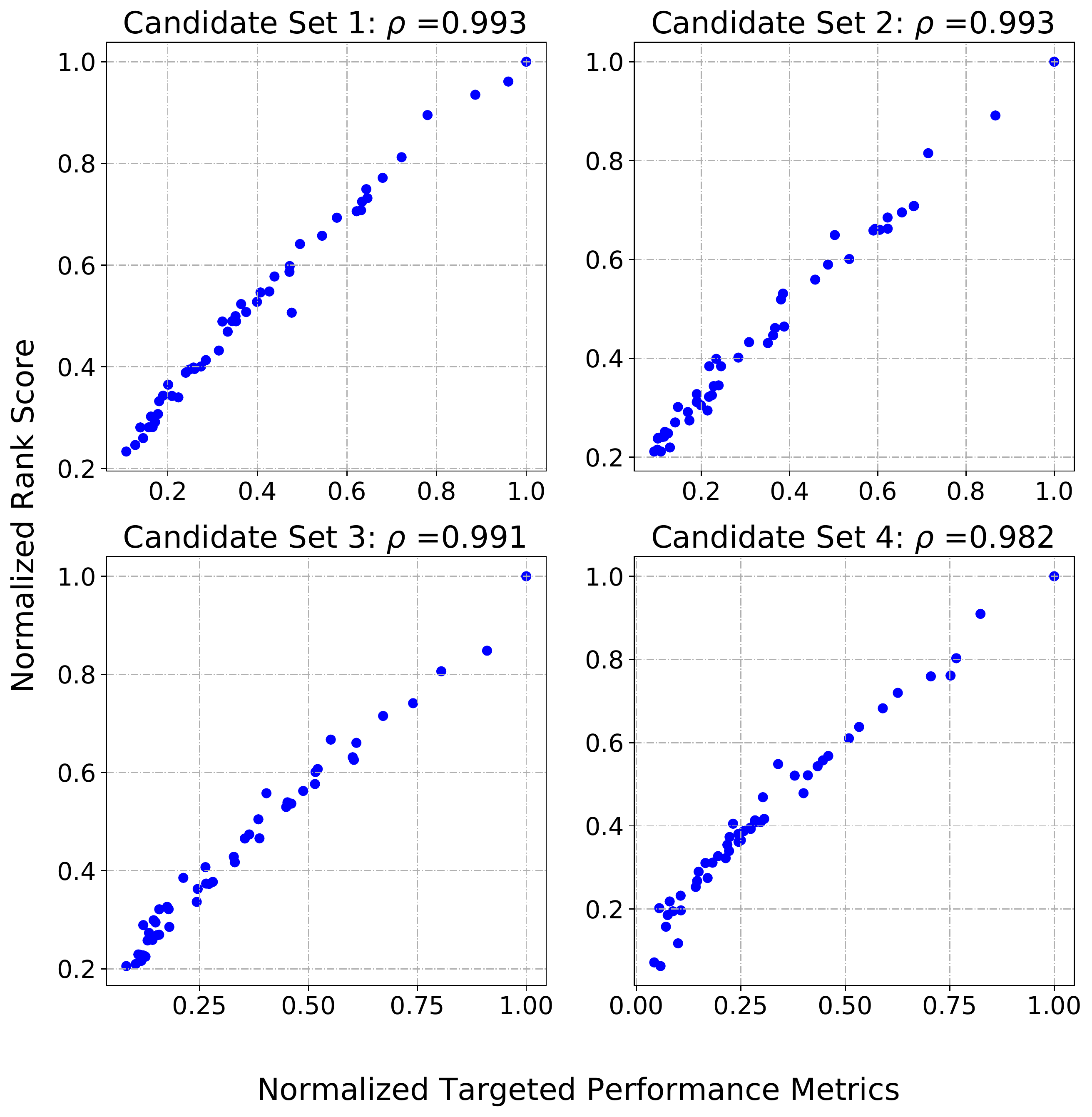}
\caption{The positive correlations between learned rank scores and the targeted performance metrics. Each blue dot represents an ad in the candidate set.}
% \vspace{-0.5em}
\label{fig:scatter_pearson}
\end{figure}

\begin{figure*}[!t]
\centering
\includegraphics[width=0.8\textwidth]{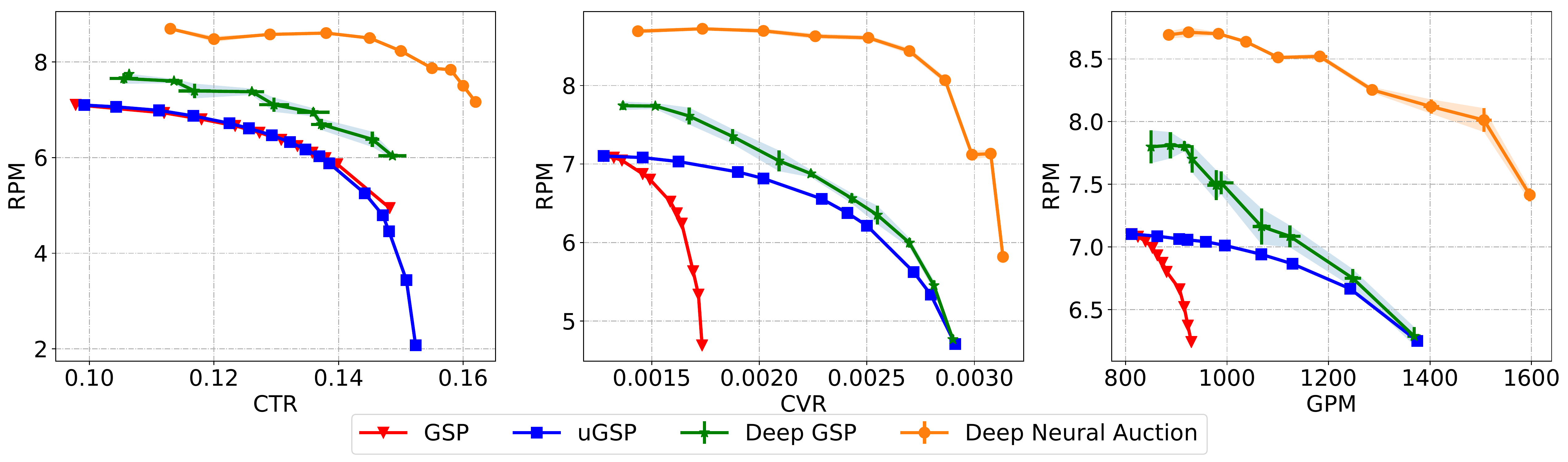}
% \vspace{-0.5em}
\caption{The performance of DNA and other baseline mechanisms in the offline experiments.}
% \vspace{-0.5em}
\label{fig:offline_exp}
\end{figure*}

We conduct experiments to compare the performance of DNA and other baseline mechanisms. In order to facilitate intuitive comparisons, we set only two performance metrics with the form $\lambda \times RPM + (1-\lambda) \times X$, where $X$ is one of the metric selected from $\{CTR, CVR, GPM\}$. For uGSP, we set the rank score function with $\lambda \times pCTR \times bid + (1 - \lambda) \times X$. For GSP, we tune the variable $\sigma$ in the interval $[0.5, 2.0]$. For both Deep GSP and DNA, we directly set the objective by selecting the values of $\lambda$ uniformly from the interval $[0, 1]$. 

We show the relation between the learned rank scores and the targeted performance metrics, and illustrate some results in Fig.~\ref{fig:scatter_pearson}.
We calculate the Pearson’s Correlation Coefficient ($\rho$), which is $0.989$ on the test data set, together with the p-value less than $1\mathrm{e}{-7}$. This result indicates the strong positive correlation between the learned rank scores and the performance metrics, implying that the ads with higher targeted objectives also have higher rank scores. 
This would encourage advertisers to optimize their ads' quality (such as CTR) to enhance their competitiveness in the auctions. 
% which has a positive effect on the ecological health of the entire auction ecosystem.

As some performance metrics may be conflicting, we next plot the Pareto Front for different baselines in Fig.~\ref{fig:offline_exp}.
% Since Deep GSP utilizes a deep model with a learning algorithm, we also illustrate its error bar to demonstrate its stability.
We observe that all the learning-based methods (both Deep GSP and DNA) are above the curves of other baselines. The flexible learning-based rank score models have the ability to perform automatic feature extraction from raw data. Learning-based methods can alleviate the problem of inaccurate predicted values (such as $pCTR$) used in GSP and uGSP to some extend, and learn ad auctions directly from the real feedback. We also note that the performance of GSP is poor when considering CVR and GPM, as GSP does not model the effect of these indicators explicitly in its rank score formulation. 

DNA outperforms Deep GSP baseline by a clear margin. The rank score function of Deep GSP is only conditioned on each ad's private information. While in DNA, it also contains a set embedding which models the context of auction from the candidate ads set explicitly, making it more competitive than the rank scores in the Deep GSP auctions. We analyze this set embedding in more details in Appendix~\ref{app:embeddings}. The upgrade of training method also contributes to this superior performance improvement. Deep GSP treats the whole process of allocation as the environment and uses exploration based algorithms (\ie, policy search) to optimize the rank score functions. In comparison, DNA directly differentiates the sorting procedure in allocation, which is more data-efficient.
% For DNA, by using differentiable sorting operation, making the sorting procedure compatible with differentiablity, it constructs a computable correlation between learned rank scores and the optimization objectives. Thus we can use the automatic end-to-end deep learning pipeline, which is more sample-efficient.

\subsubsection{Evaluation on IC Property}
\label{sec:exp_ic}
We present the IC property of DNA, \ie, the regret metric $\boldsymbol{\Psi}$, in Table~\ref{tab:ic}.
% \footnote{We empirically find that non-zero regret occurs to only a few (\eg the top-K) bidders. For a clearer illustration of the regret metric, we take average over these bidders only, instead of over all $N$ bidders.} 
We compare with only the regret of truthful bidding in the Utility-based Generalized First Price (uGFP) auction, as the regret of other mechanisms (GSP and uGSP) are all $0$. uGFP mechanism allocates the ad slots in the same way as uGSP while the payment of a winning advertiser is simply her bid.
% We compare the results with only the utility-based generalized first price (uGFP) auction mechanism, as that of other mechanisms are all 0. uGFP mechanism allocates the ad slots in the same way as uGSP while the payment of a winning advertiser is simply her bid.
One can observe from Table~\ref{tab:ic} that $\boldsymbol{\Psi}_v$ of both DNA and uGFP are $0$, which indicates that advertisers could not win higher slots under the payment constraints. For $\boldsymbol{\Psi}_p$, DNA outperforms uGFP significantly under all experiment settings. 
For example of the first row (1.0$\times$RPM), an advertiser could be undercharged at most $0.042\%$ payment in DNA with the same allocation result, which is $13.312\%$ in uGFP.
% Intuitively, for the first objective function, an advertiser could be undercharged at most $0.042\%$ payment in DNA with the same allocation result, which is $13.312\%$ in uGFP.

% \footnote{These values are averaged over 50 advertisers, while the regret may occur for only a few advertisers. If we take average over these bidders only, the values of $\boldsymbol{\Psi}_p$ are 0.0478\% and 16.78\% respectively.}

%More importantly, the values of $\boldsymbol{\Psi}_p$ of DNA are quite close to 0, that is, advertisers could hardly manipulate their bids to be undercharged maintaining the current ad slots. This result suggests that DNA guarantees a good performance on IC property.
%and the approximation mentioned in Section~\ref{sec} do not 

\begin{table}[!t]
    % \caption{The comparison on the incentive compatibility metric between DNA and other baseline mechanisms.}
    \caption{IC Metric ($\boldsymbol{\Psi}$) experiments under four tasks with 1000 PV requests randomly selected from the test data. For each bidder, we randomly generate 100 perturbations (ranging from $0.0$ to $2.0$ times) to her value.}
    %\vspace{-0.6em}
    \small
    \centering
\begin{tabular}{|c|c|c|c|c|}
\hline
\multirow{2}*{} & \multicolumn{2}{c|}{DNA}                      & \multicolumn{2}{c|}{uGFP}                      \\ \cline{2-5} 
                  & $\boldsymbol{\Psi}_v$ & $\boldsymbol{\Psi}_p$ & $\boldsymbol{\Psi}_v$ & $\boldsymbol{\Psi}_p$ \\ \hline
1.0$\times$RPM                & $0$  & \textbf{0.042\%}  & $0$  & $13.312\%$                \\ \hline
0.5$\times$RPM+0.5$\times$CTR & $0$  & \textbf{0.059\%}  & $0$  & $21.616\%$                \\ \hline
0.5$\times$RPM+0.5$\times$CVR & $0$  & \textbf{0.118\%}  & $0$  & $19.280\%$                \\ \hline
0.5$\times$RPM+0.5$\times$GPM & $0$  & \textbf{0.028\%}  & $0$  & $16.400\%$                \\ \hline
% AVG                           & $0$  & $0.061\%$  & $0$  & $17.652\%$                \\ \hline
\end{tabular}
\label{tab:ic}
    %\vspace{-0.6em}
\end{table}

\subsection{Online Experiments}
We present the online experiments by deploying the proposed DNA in \emph{Taobao} advertising system. As we only use the trained model to make inference in online system, we set the temperature parameter $\tau=0$ in the differentiable sorting engine to output the winning ads as well as the corresponding payments to the online advertising platform. Other deployment details are described in Appendix~\ref{app:deploy}.

\begin{table}[!t]
    \caption{Online A/B test compared with uGSP on promoting different performance metrics, keeping the same RPM level.}
    %\vspace{-0.6em}
    \centering
    \small
    \begin{tabular}{|c|c|c|}
        \hline
        \% Improved & Deep GSP & DNA\\
        \hline
        CTR  & +6.43\% & \textbf{+11.58\%} \\
        \hline
        CVR  & +6.38\% & \textbf{+31.26\%} \\
        \hline
        GPM  & +2.77\% & \textbf{+16.17\%} \\
        \hline
    \end{tabular}
    \label{tab:onlines}
    %\vspace{-0.6em}
\end{table}

\begin{table}[!t]
    \caption{Online A/B test (nearly two months) compared with GSP on promoting all performance metrics.}
    %\vspace{-0.6em}
    \centering
    \small
    \begin{tabular}{|c|c|c|c|c|c|c|}
        \hline
        & RPM & CTR & CVR & GPM \\
        \hline
        \% Improved & +5.68\% & +18.93\% & +14.68\% & +14.53\% \\
        \hline
    \end{tabular}
    \label{tab:onlines_longterm}
    %\vspace{-0.6em}
\end{table}

To demonstrate the performance of DNA,
% ability of DNA in optimizing the multiple performance metrics,
we conduct online A/B tests with 1\% of whole production traffic from Jan 25, 2021 to Feb 8, 2021 (about one billion auctions). We also consider RPM, CTR, CVR and GPM metrics and conduct online experiments as in the offline experiments, \ie, setting only two performance metrics at a time. In order to make fair and efficient comparisons between different baselines in production traffic, we set the $\lambda$ in uGSP with $0.8$, and tune the $\lambda$'s of both Deep GSP and DNA until the observed RPM performance reaches the same level with the one in uGSP. Then we record the relative improvements of the other metrics of Deep GSP and DNA compared with uGSP, which is shown in table~\ref{tab:onlines}. From the results, we can find that the DNA mechanism achieves the highest promotion for CTR, CVR and GPM.
To verify the performance stability of the DNA, we conduct a relatively long-term experiment (nearly two months) to compare with the GSP auction. Table~\ref{tab:onlines_longterm} shows that all the performance metrics related to users, advertisers, and advertising platform are promoted. Considering
the massive advertisers and users, the promotion of marketing performance verifies the effectiveness of our proposed DNA framework.

% Table~\ref{tab:onlines_longterm} shows the performance promotion on considering all performance metrics, compared with GSP. These results are obtained over a relatively long period (about two weeks), and demonstrate the effectiveness of our proposed DNA.
% Other online experiments are introduced in Appendix~\ref{app:online}.

\section{Related Work}

A plethora of works have used learning-based techniques for revenue-maximizing mechanism design in theoretical auction settings. 
\citeauthor{conitzer2002complexity}~\cite{conitzer2002complexity} proposed the paradigm of automated mechanism design (AMD) and laid the groundwork for this direction. Since then, many works \cite{golrezaei2017boosted, lahaie2011kernel} have adopted learning approaches for mechanism design problems. Recently,~\citeauthor{duetting2019optimal}~\cite{duetting2019optimal} first leveraged deep neural networks for the automated design of optimal auctions. Several other works extended this study for various scenarios~\cite{feng2018deep, golowich2018deep, rahme2021auction}.
% However, they mainly focused on theoretical auction settings under certain assumptions, and hence could not apply to the practical settings seamlessly.

A particular stream of research have focused on the mechanism design in online advertising. The GSP and VCG auction have been widely adopted and investigated in various advertising system~\cite{edelman2007internet, lahaie2007revenue}.
%Early studies \cite{lahaie2007revenue,feldman2011revenue, thompson2013revenue} investigated and further extended these mechanisms, \eg,~\citeauthor{thompson2013revenue}~\cite{thompson2013revenue} studied many approaches to increasing revenue under the GSP framework, such as reserve prices and exponential parameters. 
Building on the success of deep reinforcement learning,~\citeauthor{tang2017reinforcement} \emph{et al.}~\cite{tang2017reinforcement, cai2018reinforcement} proposed reinforcement mechanism design for the optimization of reserve prices in online advertising. 
% However, nearly all of the above works concentrated on single optimization metrics (\ie, revenue or social welfare), while our work incorporates multiple performance metrics in industrial e-commerce advertising besides revenue. 
A recently proposed learning-based ad auction mechanism, called Deep GSP~\cite{zhang2021optimizing}, leveraged the deep learning technique to optimize multiple performance metrics in e-commerce advertising. 

\section{Conclusion}
In this paper, we have proposed a Deep Neural Auction mechanism, towards learning data efficient and end-to-end auction mechanisms for e-commerce advertising. We have deployed the DNA mechanism on one of the leading e-commerce advertising platforms, \emph{Taobao}. The offline experiments as well as the online A/B test showed that DNA mechanism significantly outperformed other existing auction mechanism baselines on optimizing multiple performance metrics.

\begin{acks}
This work was supported in part by Science and Technology Innovation 2030 –``New Generation Artificial Intelligence'' Major Project No. 2018AAA0100905, in part by China NSF grant No. 62025204, 62072303, 61902248, and 61972254, and in part by Alibaba Group through Alibaba Innovation Research Program, and in part by Shanghai Science and Technology fund 20PJ1407900.
The authors would like to thank Rui Du, Haiping Huang, Haiyang He and Guan Wang who did the really hard work for online system implementation. 
% The authors would also like to thank all anonymous reviewers for their valuable help and suggestions.
The opinions, findings, conclusions, and recommendations expressed in this paper are those of the authors and do not necessarily reflect the views of the funding agencies or the government.
\end{acks}

\bibliographystyle{ACM-Reference-Format}
\bibliography{sample-base}


\begin{thebibliography}{41}

%%% the Web address.

\ifx \showCODEN    \undefined \def \showCODEN     #1{\unskip}     \fi
\ifx \showDOI      \undefined \def \showDOI       #1{#1}\fi
\ifx \showISBNx    \undefined \def \showISBNx     #1{\unskip}     \fi
\ifx \showISBNxiii \undefined \def \showISBNxiii  #1{\unskip}     \fi
\ifx \showISSN     \undefined \def \showISSN      #1{\unskip}     \fi
\ifx \showLCCN     \undefined \def \showLCCN      #1{\unskip}     \fi
\ifx \shownote     \undefined \def \shownote      #1{#1}          \fi
\ifx \showarticletitle \undefined \def \showarticletitle #1{#1}   \fi
\ifx \showURL      \undefined \def \showURL       {\relax}        \fi
% invisible to TeX
\providecommand\bibfield[2]{#2}
\providecommand\bibinfo[2]{#2}
\providecommand\natexlab[1]{#1}
\providecommand\showeprint[2][]{arXiv:#2}

\bibitem[\protect\citeauthoryear{Aggarwal, Muthukrishnan, P{\'a}l, and
  P{\'a}l}{Aggarwal et~al\mbox{.}}{2009}]%
        {aggarwal2009general}
\bibfield{author}{\bibinfo{person}{Gagan Aggarwal}, \bibinfo{person}{S
  Muthukrishnan}, \bibinfo{person}{D{\'a}vid P{\'a}l}, {and}
  \bibinfo{person}{Martin P{\'a}l}.} \bibinfo{year}{2009}\natexlab{}.
\newblock \showarticletitle{General auction mechanism for search advertising}.
  In \bibinfo{booktitle}{\emph{WWW}}. \bibinfo{pages}{241--250}.
\newblock


\bibitem[\protect\citeauthoryear{Bachrach, Ceppi, Kash, Key, and
  Kurokawa}{Bachrach et~al\mbox{.}}{2014}]%
        {bachrach2014optimising}
\bibfield{author}{\bibinfo{person}{Yoram Bachrach}, \bibinfo{person}{Sofia
  Ceppi}, \bibinfo{person}{Ian~A Kash}, \bibinfo{person}{Peter Key}, {and}
  \bibinfo{person}{David Kurokawa}.} \bibinfo{year}{2014}\natexlab{}.
\newblock \showarticletitle{Optimising trade-offs among stakeholders in ad
  auctions}. In \bibinfo{booktitle}{\emph{EC}}. \bibinfo{pages}{75--92}.
\newblock


\bibitem[\protect\citeauthoryear{Battaglia, Hamrick, Bapst, Sanchez-Gonzalez,
  Zambaldi, Malinowski, Tacchetti, Raposo, Santoro, Faulkner,
  et~al\mbox{.}}{Battaglia et~al\mbox{.}}{2018}]%
        {battaglia2018relational}
\bibfield{author}{\bibinfo{person}{Peter~W Battaglia},
  \bibinfo{person}{Jessica~B Hamrick}, \bibinfo{person}{Victor Bapst},
  \bibinfo{person}{Alvaro Sanchez-Gonzalez}, \bibinfo{person}{Vinicius
  Zambaldi}, \bibinfo{person}{Mateusz Malinowski}, \bibinfo{person}{Andrea
  Tacchetti}, \bibinfo{person}{David Raposo}, \bibinfo{person}{Adam Santoro},
  \bibinfo{person}{Ryan Faulkner}, {et~al\mbox{.}}}
  \bibinfo{year}{2018}\natexlab{}.
\newblock \showarticletitle{Relational inductive biases, deep learning, and
  graph networks}.
\newblock \bibinfo{journal}{\emph{arXiv preprint arXiv:1806.01261}}
  (\bibinfo{year}{2018}).
\newblock


\bibitem[\protect\citeauthoryear{Borisov, Kiseleva, Markov, and
  de~Rijke}{Borisov et~al\mbox{.}}{2018}]%
        {borisov2018calibration}
\bibfield{author}{\bibinfo{person}{Alexey Borisov}, \bibinfo{person}{Julia
  Kiseleva}, \bibinfo{person}{Ilya Markov}, {and} \bibinfo{person}{Maarten de
  Rijke}.} \bibinfo{year}{2018}\natexlab{}.
\newblock \showarticletitle{Calibration: A simple way to improve click models}.
  In \bibinfo{booktitle}{\emph{CIKM}}. \bibinfo{pages}{1503--1506}.
\newblock


\bibitem[\protect\citeauthoryear{Cai, Filos-Ratsikas, Tang, and Zhang}{Cai
  et~al\mbox{.}}{2018}]%
        {cai2018reinforcement}
\bibfield{author}{\bibinfo{person}{Qingpeng Cai}, \bibinfo{person}{Aris
  Filos-Ratsikas}, \bibinfo{person}{Pingzhong Tang}, {and}
  \bibinfo{person}{Yiwei Zhang}.} \bibinfo{year}{2018}\natexlab{}.
\newblock \showarticletitle{Reinforcement Mechanism Design for e-commerce}. In
  \bibinfo{booktitle}{\emph{WWW}}. \bibinfo{pages}{1339--1348}.
\newblock


\bibitem[\protect\citeauthoryear{Cheng, Koc, Harmsen, Shaked, Chandra, Aradhye,
  Anderson, Corrado, Chai, Ispir, et~al\mbox{.}}{Cheng et~al\mbox{.}}{2016}]%
        {cheng2016wide}
\bibfield{author}{\bibinfo{person}{Heng-Tze Cheng}, \bibinfo{person}{Levent
  Koc}, \bibinfo{person}{Jeremiah Harmsen}, \bibinfo{person}{Tal Shaked},
  \bibinfo{person}{Tushar Chandra}, \bibinfo{person}{Hrishi Aradhye},
  \bibinfo{person}{Glen Anderson}, \bibinfo{person}{Greg Corrado},
  \bibinfo{person}{Wei Chai}, \bibinfo{person}{Mustafa Ispir}, {et~al\mbox{.}}}
  \bibinfo{year}{2016}\natexlab{}.
\newblock \showarticletitle{Wide \& deep learning for recommender systems}. In
  \bibinfo{booktitle}{\emph{Proceedings of the 1st workshop on deep learning
  for recommender systems}}. \bibinfo{pages}{7--10}.
\newblock


\bibitem[\protect\citeauthoryear{Clevert, Unterthiner, and Hochreiter}{Clevert
  et~al\mbox{.}}{2016}]%
        {DBLP:journals/corr/ClevertUH15}
\bibfield{author}{\bibinfo{person}{Djork{-}Arn{\'{e}} Clevert},
  \bibinfo{person}{Thomas Unterthiner}, {and} \bibinfo{person}{Sepp
  Hochreiter}.} \bibinfo{year}{2016}\natexlab{}.
\newblock \showarticletitle{Fast and Accurate Deep Network Learning by
  Exponential Linear Units (ELUs)}. In \bibinfo{booktitle}{\emph{ICLR}}.
\newblock


\bibitem[\protect\citeauthoryear{Conitzer and Sandholm}{Conitzer and
  Sandholm}{2002}]%
        {conitzer2002complexity}
\bibfield{author}{\bibinfo{person}{Vincent Conitzer} {and}
  \bibinfo{person}{Tuomas Sandholm}.} \bibinfo{year}{2002}\natexlab{}.
\newblock \showarticletitle{Complexity of mechanism design}. In
  \bibinfo{booktitle}{\emph{UAI}}. \bibinfo{pages}{103--110}.
\newblock


\bibitem[\protect\citeauthoryear{Daniels and Velikova}{Daniels and
  Velikova}{2010}]%
        {daniels2010monotone}
\bibfield{author}{\bibinfo{person}{Hennie Daniels} {and}
  \bibinfo{person}{Marina Velikova}.} \bibinfo{year}{2010}\natexlab{}.
\newblock \showarticletitle{Monotone and partially monotone neural networks}.
\newblock \bibinfo{journal}{\emph{IEEE Transactions on Neural Networks}}
  \bibinfo{volume}{21}, \bibinfo{number}{6} (\bibinfo{year}{2010}),
  \bibinfo{pages}{906--917}.
\newblock


\bibitem[\protect\citeauthoryear{de~Avila Belbute-Peres, Smith, Allen,
  Tenenbaum, and Kolter}{de~Avila Belbute-Peres et~al\mbox{.}}{2018}]%
        {de2018end}
\bibfield{author}{\bibinfo{person}{Filipe de Avila Belbute-Peres},
  \bibinfo{person}{Kevin Smith}, \bibinfo{person}{Kelsey Allen},
  \bibinfo{person}{Josh Tenenbaum}, {and} \bibinfo{person}{J~Zico Kolter}.}
  \bibinfo{year}{2018}\natexlab{}.
\newblock \showarticletitle{End-to-end differentiable physics for learning and
  control}.
\newblock \bibinfo{journal}{\emph{NeurIPS}}  \bibinfo{volume}{31}
  (\bibinfo{year}{2018}), \bibinfo{pages}{7178--7189}.
\newblock


\bibitem[\protect\citeauthoryear{Deng, Wang, Tan, Xu, and Gai}{Deng
  et~al\mbox{.}}{2021}]%
        {deng2021calibrating}
\bibfield{author}{\bibinfo{person}{Chao Deng}, \bibinfo{person}{Hao Wang},
  \bibinfo{person}{Qing Tan}, \bibinfo{person}{Jian Xu}, {and}
  \bibinfo{person}{Kun Gai}.} \bibinfo{year}{2021}\natexlab{}.
\newblock \showarticletitle{Calibrating User Response Predictions in Online
  Advertising}. In \bibinfo{booktitle}{\emph{ECML PKDD 2020}}.
  \bibinfo{pages}{208--223}.
\newblock


\bibitem[\protect\citeauthoryear{D{\"u}tting, Feng, Narasimhan, Parkes, and
  Ravindranath}{D{\"u}tting et~al\mbox{.}}{2019}]%
        {duetting2019optimal}
\bibfield{author}{\bibinfo{person}{Paul D{\"u}tting}, \bibinfo{person}{Zhe
  Feng}, \bibinfo{person}{Harikrishna Narasimhan}, \bibinfo{person}{David
  Parkes}, {and} \bibinfo{person}{Sai~Srivatsa Ravindranath}.}
  \bibinfo{year}{2019}\natexlab{}.
\newblock \showarticletitle{Optimal Auctions through Deep Learning}. In
  \bibinfo{booktitle}{\emph{ICML}}. \bibinfo{pages}{1706--1715}.
\newblock


\bibitem[\protect\citeauthoryear{Edelman and Ostrovsky}{Edelman and
  Ostrovsky}{2007}]%
        {edelman2007strategic}
\bibfield{author}{\bibinfo{person}{Benjamin Edelman} {and}
  \bibinfo{person}{Michael Ostrovsky}.} \bibinfo{year}{2007}\natexlab{}.
\newblock \showarticletitle{Strategic bidder behavior in sponsored search
  auctions}.
\newblock \bibinfo{journal}{\emph{Decision support systems}}
  \bibinfo{volume}{43}, \bibinfo{number}{1} (\bibinfo{year}{2007}),
  \bibinfo{pages}{192--198}.
\newblock


\bibitem[\protect\citeauthoryear{Edelman, Ostrovsky, and Schwarz}{Edelman
  et~al\mbox{.}}{2007}]%
        {edelman2007internet}
\bibfield{author}{\bibinfo{person}{Benjamin Edelman}, \bibinfo{person}{Michael
  Ostrovsky}, {and} \bibinfo{person}{Michael Schwarz}.}
  \bibinfo{year}{2007}\natexlab{}.
\newblock \showarticletitle{Internet advertising and the generalized
  second-price auction: Selling billions of dollars worth of keywords}.
\newblock \bibinfo{journal}{\emph{American economic review}}
  \bibinfo{volume}{97}, \bibinfo{number}{1} (\bibinfo{year}{2007}),
  \bibinfo{pages}{242--259}.
\newblock


\bibitem[\protect\citeauthoryear{Edwards and Storkey}{Edwards and
  Storkey}{2017}]%
        {DBLP:conf/iclr/EdwardsS17}
\bibfield{author}{\bibinfo{person}{Harrison Edwards} {and}
  \bibinfo{person}{Amos~J. Storkey}.} \bibinfo{year}{2017}\natexlab{}.
\newblock \showarticletitle{Towards a Neural Statistician}. In
  \bibinfo{booktitle}{\emph{ICLR}}.
\newblock


\bibitem[\protect\citeauthoryear{Feng, Narasimhan, and Parkes}{Feng
  et~al\mbox{.}}{2018}]%
        {feng2018deep}
\bibfield{author}{\bibinfo{person}{Zhe Feng}, \bibinfo{person}{Harikrishna
  Narasimhan}, {and} \bibinfo{person}{David~C Parkes}.}
  \bibinfo{year}{2018}\natexlab{}.
\newblock \showarticletitle{Deep learning for revenue-optimal auctions with
  budgets}. In \bibinfo{booktitle}{\emph{AAMAS}}. \bibinfo{pages}{354--362}.
\newblock


\bibitem[\protect\citeauthoryear{Feng, Schrijvers, and Sodomka}{Feng
  et~al\mbox{.}}{2019}]%
        {feng2019online}
\bibfield{author}{\bibinfo{person}{Zhe Feng}, \bibinfo{person}{Okke
  Schrijvers}, {and} \bibinfo{person}{Eric Sodomka}.}
  \bibinfo{year}{2019}\natexlab{}.
\newblock \showarticletitle{Online learning for measuring incentive
  compatibility in ad auctions}. In \bibinfo{booktitle}{\emph{WWW}}.
  \bibinfo{pages}{2729--2735}.
\newblock


\bibitem[\protect\citeauthoryear{Goldfarb and Tucker}{Goldfarb and
  Tucker}{2011}]%
        {goldfarb2011online}
\bibfield{author}{\bibinfo{person}{Avi Goldfarb} {and}
  \bibinfo{person}{Catherine Tucker}.} \bibinfo{year}{2011}\natexlab{}.
\newblock \showarticletitle{Online display advertising: Targeting and
  obtrusiveness}.
\newblock \bibinfo{journal}{\emph{Marketing Science}} \bibinfo{volume}{30},
  \bibinfo{number}{3} (\bibinfo{year}{2011}), \bibinfo{pages}{389--404}.
\newblock


\bibitem[\protect\citeauthoryear{Golowich, Narasimhan, and Parkes}{Golowich
  et~al\mbox{.}}{2018}]%
        {golowich2018deep}
\bibfield{author}{\bibinfo{person}{Noah Golowich}, \bibinfo{person}{Harikrishna
  Narasimhan}, {and} \bibinfo{person}{David~C Parkes}.}
  \bibinfo{year}{2018}\natexlab{}.
\newblock \showarticletitle{Deep Learning for Multi-Facility Location Mechanism
  Design.}. In \bibinfo{booktitle}{\emph{IJCAI}}. \bibinfo{pages}{261--267}.
\newblock


\bibitem[\protect\citeauthoryear{Golrezaei, Lin, Mirrokni, and
  Nazerzadeh}{Golrezaei et~al\mbox{.}}{2017}]%
        {golrezaei2017boosted}
\bibfield{author}{\bibinfo{person}{Negin Golrezaei}, \bibinfo{person}{Max Lin},
  \bibinfo{person}{Vahab Mirrokni}, {and} \bibinfo{person}{Hamid Nazerzadeh}.}
  \bibinfo{year}{2017}\natexlab{}.
\newblock \showarticletitle{Boosted second price auctions: Revenue optimization
  for heterogeneous bidders}.
\newblock \bibinfo{journal}{\emph{Available at SSRN 3016465}}
  (\bibinfo{year}{2017}).
\newblock


\bibitem[\protect\citeauthoryear{Grover, Wang, Zweig, and Ermon}{Grover
  et~al\mbox{.}}{2019}]%
        {DBLP:conf/iclr/GroverWZE19}
\bibfield{author}{\bibinfo{person}{Aditya Grover}, \bibinfo{person}{Eric Wang},
  \bibinfo{person}{Aaron Zweig}, {and} \bibinfo{person}{Stefano Ermon}.}
  \bibinfo{year}{2019}\natexlab{}.
\newblock \showarticletitle{Stochastic Optimization of Sorting Networks via
  Continuous Relaxations}. In \bibinfo{booktitle}{\emph{ICLR}}.
\newblock


\bibitem[\protect\citeauthoryear{Ha and Schmidhuber}{Ha and
  Schmidhuber}{2018}]%
        {ha2018world}
\bibfield{author}{\bibinfo{person}{David Ha} {and} \bibinfo{person}{J{\"u}rgen
  Schmidhuber}.} \bibinfo{year}{2018}\natexlab{}.
\newblock \showarticletitle{World models}.
\newblock \bibinfo{journal}{\emph{arXiv preprint arXiv:1803.10122}}
  (\bibinfo{year}{2018}).
\newblock


\bibitem[\protect\citeauthoryear{H{\"u}llermeier and Beringer}{H{\"u}llermeier
  and Beringer}{2006}]%
        {hullermeier2006learning}
\bibfield{author}{\bibinfo{person}{Eyke H{\"u}llermeier} {and}
  \bibinfo{person}{J{\"u}rgen Beringer}.} \bibinfo{year}{2006}\natexlab{}.
\newblock \showarticletitle{Learning from ambiguously labeled examples}.
\newblock \bibinfo{journal}{\emph{Intelligent Data Analysis}}
  \bibinfo{volume}{10}, \bibinfo{number}{5} (\bibinfo{year}{2006}),
  \bibinfo{pages}{419--439}.
\newblock


\bibitem[\protect\citeauthoryear{Kurutach, Clavera, Duan, Tamar, and
  Abbeel}{Kurutach et~al\mbox{.}}{2018}]%
        {kurutach2018modelensemble}
\bibfield{author}{\bibinfo{person}{Thanard Kurutach}, \bibinfo{person}{Ignasi
  Clavera}, \bibinfo{person}{Yan Duan}, \bibinfo{person}{Aviv Tamar}, {and}
  \bibinfo{person}{Pieter Abbeel}.} \bibinfo{year}{2018}\natexlab{}.
\newblock \showarticletitle{Model-Ensemble Trust-Region Policy Optimization}.
  In \bibinfo{booktitle}{\emph{ICLR}}.
\newblock


\bibitem[\protect\citeauthoryear{Lahaie}{Lahaie}{2011}]%
        {lahaie2011kernel}
\bibfield{author}{\bibinfo{person}{S{\'e}bastien Lahaie}.}
  \bibinfo{year}{2011}\natexlab{}.
\newblock \showarticletitle{A kernel-based iterative combinatorial auction}. In
  \bibinfo{booktitle}{\emph{AAAI}}, Vol.~\bibinfo{volume}{25}.
\newblock


\bibitem[\protect\citeauthoryear{Lahaie and Pennock}{Lahaie and
  Pennock}{2007}]%
        {lahaie2007revenue}
\bibfield{author}{\bibinfo{person}{S{\'e}bastien Lahaie} {and}
  \bibinfo{person}{David~M Pennock}.} \bibinfo{year}{2007}\natexlab{}.
\newblock \showarticletitle{Revenue analysis of a family of ranking rules for
  keyword auctions}. In \bibinfo{booktitle}{\emph{EC}}.
  \bibinfo{pages}{50--56}.
\newblock


\bibitem[\protect\citeauthoryear{Moerland, Broekens, and Jonker}{Moerland
  et~al\mbox{.}}{2020}]%
        {moerland2020model}
\bibfield{author}{\bibinfo{person}{Thomas~M Moerland}, \bibinfo{person}{Joost
  Broekens}, {and} \bibinfo{person}{Catholijn~M Jonker}.}
  \bibinfo{year}{2020}\natexlab{}.
\newblock \showarticletitle{Model-based reinforcement learning: A survey}.
\newblock \bibinfo{journal}{\emph{arXiv preprint arXiv:2006.16712}}
  (\bibinfo{year}{2020}).
\newblock


\bibitem[\protect\citeauthoryear{Myerson}{Myerson}{1981}]%
        {myerson1981optimal}
\bibfield{author}{\bibinfo{person}{Roger~B Myerson}.}
  \bibinfo{year}{1981}\natexlab{}.
\newblock \showarticletitle{Optimal auction design}.
\newblock \bibinfo{journal}{\emph{Mathematics of operations research}}
  \bibinfo{volume}{6}, \bibinfo{number}{1} (\bibinfo{year}{1981}),
  \bibinfo{pages}{58--73}.
\newblock


\bibitem[\protect\citeauthoryear{Rahme, Jelassi, and Weinberg}{Rahme
  et~al\mbox{.}}{2021}]%
        {rahme2021auction}
\bibfield{author}{\bibinfo{person}{Jad Rahme}, \bibinfo{person}{Samy Jelassi},
  {and} \bibinfo{person}{S.~Matthew Weinberg}.}
  \bibinfo{year}{2021}\natexlab{}.
\newblock \showarticletitle{Auction Learning as a Two-Player Game}. In
  \bibinfo{booktitle}{\emph{ICLR}}.
\newblock


\bibitem[\protect\citeauthoryear{Shen, Tang, and Zuo}{Shen
  et~al\mbox{.}}{2019}]%
        {shen2019automated}
\bibfield{author}{\bibinfo{person}{Weiran Shen}, \bibinfo{person}{Pingzhong
  Tang}, {and} \bibinfo{person}{Song Zuo}.} \bibinfo{year}{2019}\natexlab{}.
\newblock \showarticletitle{Automated mechanism design via neural networks}. In
  \bibinfo{booktitle}{\emph{AAMAS}}. \bibinfo{pages}{215--223}.
\newblock


\bibitem[\protect\citeauthoryear{Tang}{Tang}{2017}]%
        {tang2017reinforcement}
\bibfield{author}{\bibinfo{person}{Pingzhong Tang}.}
  \bibinfo{year}{2017}\natexlab{}.
\newblock \showarticletitle{Reinforcement mechanism design.}. In
  \bibinfo{booktitle}{\emph{IJCAI}}. \bibinfo{pages}{5146--5150}.
\newblock


\bibitem[\protect\citeauthoryear{Van~der Maaten and Hinton}{Van~der Maaten and
  Hinton}{2008}]%
        {van2008visualizing}
\bibfield{author}{\bibinfo{person}{Laurens Van~der Maaten} {and}
  \bibinfo{person}{Geoffrey Hinton}.} \bibinfo{year}{2008}\natexlab{}.
\newblock \showarticletitle{Visualizing data using t-SNE.}
\newblock \bibinfo{journal}{\emph{Journal of machine learning research}}
  \bibinfo{volume}{9}, \bibinfo{number}{11} (\bibinfo{year}{2008}).
\newblock


\bibitem[\protect\citeauthoryear{Varian}{Varian}{2007}]%
        {varian2007position}
\bibfield{author}{\bibinfo{person}{Hal~R Varian}.}
  \bibinfo{year}{2007}\natexlab{}.
\newblock \showarticletitle{Position auctions}.
\newblock \bibinfo{journal}{\emph{international Journal of industrial
  Organization}} \bibinfo{volume}{25}, \bibinfo{number}{6}
  (\bibinfo{year}{2007}), \bibinfo{pages}{1163--1178}.
\newblock


\bibitem[\protect\citeauthoryear{Vickrey}{Vickrey}{1961}]%
        {vickrey1961counterspeculation}
\bibfield{author}{\bibinfo{person}{William Vickrey}.}
  \bibinfo{year}{1961}\natexlab{}.
\newblock \showarticletitle{Counterspeculation, auctions, and competitive
  sealed tenders}.
\newblock \bibinfo{journal}{\emph{The Journal of finance}}
  \bibinfo{volume}{16}, \bibinfo{number}{1} (\bibinfo{year}{1961}),
  \bibinfo{pages}{8--37}.
\newblock


\bibitem[\protect\citeauthoryear{Wilkens, Cavallo, and Niazadeh}{Wilkens
  et~al\mbox{.}}{2017}]%
        {wilkens2017gsp}
\bibfield{author}{\bibinfo{person}{Christopher~A Wilkens},
  \bibinfo{person}{Ruggiero Cavallo}, {and} \bibinfo{person}{Rad Niazadeh}.}
  \bibinfo{year}{2017}\natexlab{}.
\newblock \showarticletitle{GSP: the cinderella of mechanism design}. In
  \bibinfo{booktitle}{\emph{WWW}}. \bibinfo{pages}{25--32}.
\newblock


\bibitem[\protect\citeauthoryear{Yang, Li, Wang, Wu, Tan, Xu, and Gai}{Yang
  et~al\mbox{.}}{2019}]%
        {yang2019bid}
\bibfield{author}{\bibinfo{person}{Xun Yang}, \bibinfo{person}{Yasong Li},
  \bibinfo{person}{Hao Wang}, \bibinfo{person}{Di Wu}, \bibinfo{person}{Qing
  Tan}, \bibinfo{person}{Jian Xu}, {and} \bibinfo{person}{Kun Gai}.}
  \bibinfo{year}{2019}\natexlab{}.
\newblock \showarticletitle{Bid optimization by multivariable control in
  display advertising}. In \bibinfo{booktitle}{\emph{KDD}}.
  \bibinfo{pages}{1966--1974}.
\newblock


\bibitem[\protect\citeauthoryear{Zaheer, Kottur, Ravanbakhsh, Poczos,
  Salakhutdinov, and Smola}{Zaheer et~al\mbox{.}}{2017}]%
        {zaheer2017deep}
\bibfield{author}{\bibinfo{person}{Manzil Zaheer}, \bibinfo{person}{Satwik
  Kottur}, \bibinfo{person}{Siamak Ravanbakhsh}, \bibinfo{person}{Barnabas
  Poczos}, \bibinfo{person}{Russ~R Salakhutdinov}, {and}
  \bibinfo{person}{Alexander~J Smola}.} \bibinfo{year}{2017}\natexlab{}.
\newblock \showarticletitle{Deep sets}. In \bibinfo{booktitle}{\emph{NIPS}}.
  \bibinfo{pages}{3391--3401}.
\newblock


\bibitem[\protect\citeauthoryear{Zhang, Hare, and Pr{\"u}gel-Bennett}{Zhang
  et~al\mbox{.}}{2019}]%
        {zhang2019fspool}
\bibfield{author}{\bibinfo{person}{Yan Zhang}, \bibinfo{person}{Jonathon Hare},
  {and} \bibinfo{person}{Adam Pr{\"u}gel-Bennett}.}
  \bibinfo{year}{2019}\natexlab{}.
\newblock \showarticletitle{FSPool: Learning Set Representations with
  Featurewise Sort Pooling}. In \bibinfo{booktitle}{\emph{ICLR}}.
\newblock


\bibitem[\protect\citeauthoryear{Zhang, Liu, Zheng, Zhang, Xu, Pan, Yu, Wu, Xu,
  and Gai}{Zhang et~al\mbox{.}}{2021}]%
        {zhang2021optimizing}
\bibfield{author}{\bibinfo{person}{Zhilin Zhang}, \bibinfo{person}{Xiangyu
  Liu}, \bibinfo{person}{Zhenzhe Zheng}, \bibinfo{person}{Chenrui Zhang},
  \bibinfo{person}{Miao Xu}, \bibinfo{person}{Junwei Pan},
  \bibinfo{person}{Chuan Yu}, \bibinfo{person}{Fan Wu}, \bibinfo{person}{Jian
  Xu}, {and} \bibinfo{person}{Kun Gai}.} \bibinfo{year}{2021}\natexlab{}.
\newblock \showarticletitle{Optimizing Multiple Performance Metrics with Deep
  GSP Auctions for E-commerce Advertising}. In
  \bibinfo{booktitle}{\emph{WSDM}}. \bibinfo{pages}{993--1001}.
\newblock


\bibitem[\protect\citeauthoryear{Zhou, Zhu, Song, Fan, Zhu, Ma, Yan, Jin, Li,
  and Gai}{Zhou et~al\mbox{.}}{2018}]%
        {zhou2018deep}
\bibfield{author}{\bibinfo{person}{Guorui Zhou}, \bibinfo{person}{Xiaoqiang
  Zhu}, \bibinfo{person}{Chenru Song}, \bibinfo{person}{Ying Fan},
  \bibinfo{person}{Han Zhu}, \bibinfo{person}{Xiao Ma},
  \bibinfo{person}{Yanghui Yan}, \bibinfo{person}{Junqi Jin},
  \bibinfo{person}{Han Li}, {and} \bibinfo{person}{Kun Gai}.}
  \bibinfo{year}{2018}\natexlab{}.
\newblock \showarticletitle{Deep interest network for click-through rate
  prediction}. In \bibinfo{booktitle}{\emph{KDD}}. \bibinfo{pages}{1059--1068}.
\newblock


\bibitem[\protect\citeauthoryear{Zhu, Jin, Tan, Pan, Zeng, Li, and Gai}{Zhu
  et~al\mbox{.}}{2017}]%
        {zhu2017optimized}
\bibfield{author}{\bibinfo{person}{Han Zhu}, \bibinfo{person}{Junqi Jin},
  \bibinfo{person}{Chang Tan}, \bibinfo{person}{Fei Pan},
  \bibinfo{person}{Yifan Zeng}, \bibinfo{person}{Han Li}, {and}
  \bibinfo{person}{Kun Gai}.} \bibinfo{year}{2017}\natexlab{}.
\newblock \showarticletitle{Optimized cost per click in taobao display
  advertising}. In \bibinfo{booktitle}{\emph{KDD}}.
  \bibinfo{pages}{2191--2200}.
\newblock


\end{thebibliography}

\newpage
\appendix

\section{Model Configurations and Offline Training Procedure}
\label{app:training}
In the set encoder module (Fig.~\ref{fig:deepset}), $\phi_1$ consists of two fully connected layers with 128, 32 neurons respectively. $\phi_2$ is a single fully connected layer with 16 neurons, thus the final size of the set embeddings $h_i'$ is 16. The ELU non-linearity is applied to the output of every layer. In the context-aware rank score module (Fig.~\ref{fig:minmax}), we use 5 groups of 20 linear functions in the partially monotone MIN-MAX neural network, \ie, $Q=5, Z=20$.

We use the Adam optimizer with $\beta_1=0.9, \beta_2=0.99, \epsilon=1e{-8}$. Each batch contains 128 PV requests in total. We also leverage some temperature annealing schedules for adjusting $\tau$ in the differentiable sorting engine during the training process, such as polynomial decay and exponential decay. But we did not observe significant performance differences between these schedules. The offline training procedure of DNA is as follows:

 \begin{algorithm}
 \caption{Offline Training Procedure of DNA} \label{alg:alg_offline}
 \begin{algorithmic}[1]
 \REQUIRE Online log data with user behaviors, temperature annealing schedule in the differentiable sorting engine
 \STATE Data preprocessing: feature construction, ground-truth label generation
 \STATE Initialize the neural network parameters of the set encoder and the context-aware rank score function, initialize the temperature $\tau$ in the differentiable sorting engine
%  \STATE Bulid DNA network $\theta$ with deepSet(Eq.~(\ref{eq:deepset2})), MIN-MAX minmax architecture (Eq.~(\ref{eq:minmax_forward})) and differentiable sorting engine (Eq.~(\ref{eq:sort2}));
%  \STATE Initialize DNA network with $\theta$;
 \WHILE{not converged}
   \STATE Sample a random minibatch from training data
   \STATE Compute the target loss $\mathcal{L}_{tgt}$ and the cross-entropy loss $\mathcal{L}_{ce}$ with Eq.~(\ref{eq:loss1})(\ref{eq:loss2}), given the generated rank scores and payments with Eq.~(\ref{eq:deepset1})(\ref{eq:deepset2})(\ref{eq:minmax_forward})(\ref{eq:minmax_inv})
   \STATE Update the network parameters using stochastic gradient descent optimizer (\ie, Adam)
   \STATE Decrease $\tau$ by one step
 \ENDWHILE
 \end{algorithmic}
 \end{algorithm}

\section{Analysis of the Set Embeddings}
\label{app:embeddings}
\begin{figure}[!t]
\centering
\includegraphics[width=0.35\textwidth]{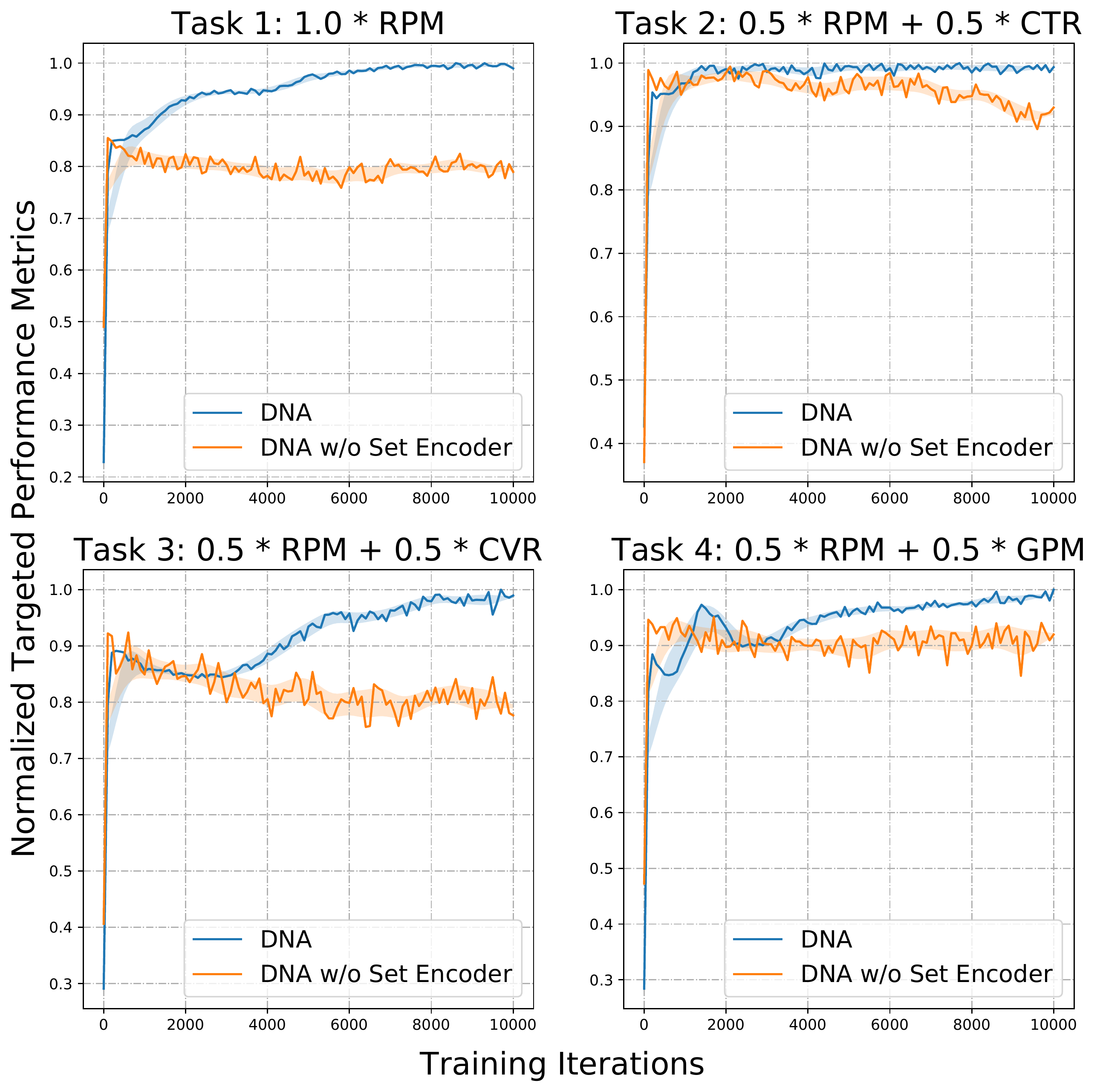}
\caption{Learning curves on DNA and DNA w/o set encoder in four training tasks.}
% \vspace{-0.5em}
\label{fig:set_comparison}
\end{figure}

\begin{figure}[!t]
\centering
\includegraphics[width=0.35\textwidth]{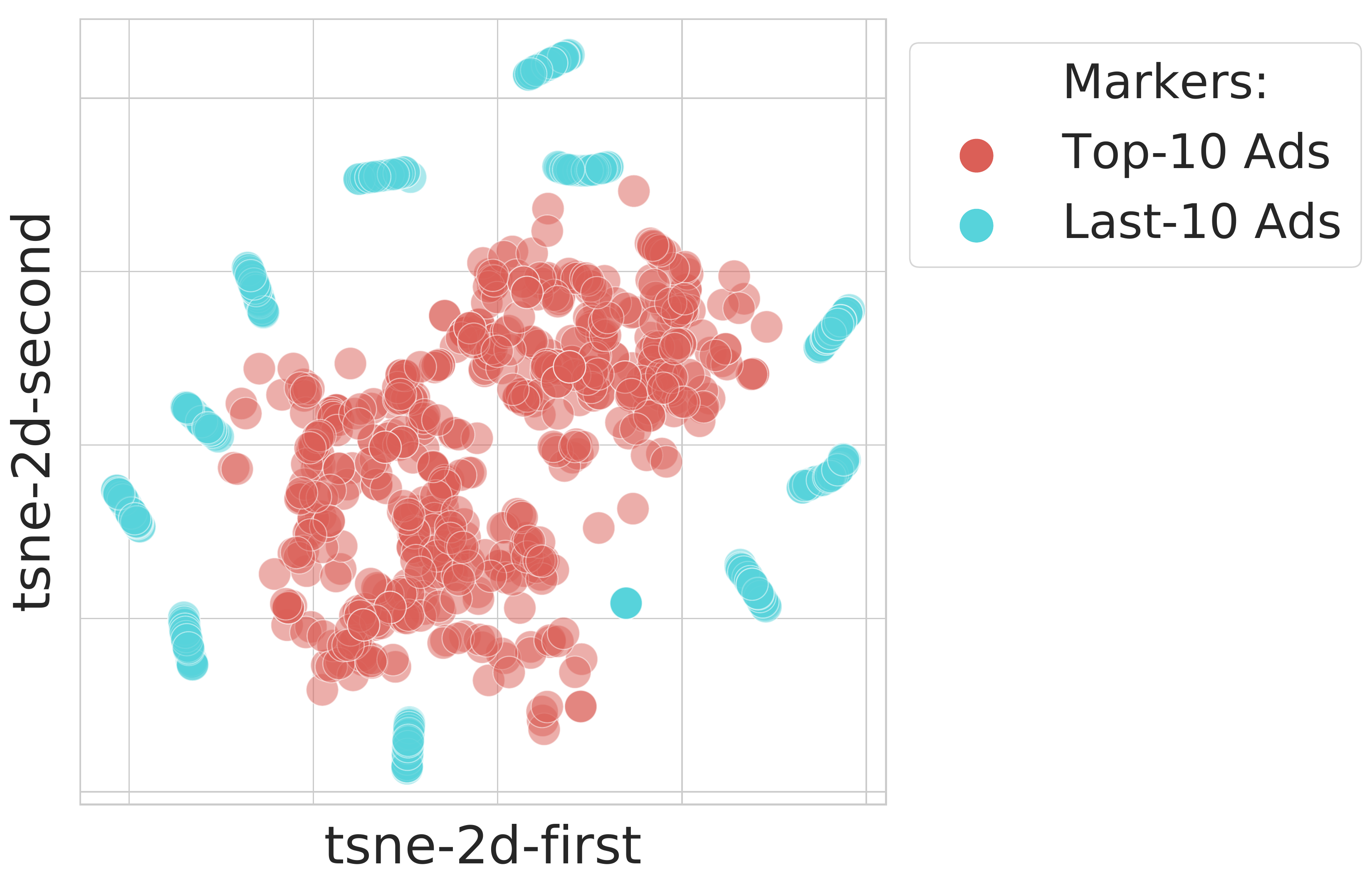}
\caption{Distribution of the set embedding for sampled tok-10 and end-10 ads in latent space using t-SNE.}
% \vspace{-0.5em}
\label{fig:tsne_plot}
\end{figure}

We next provide empirical evidence suggesting the meaningfulness of set embedding $h'_i$ learned by the set encoder. We conducted experiments on training DNA mechanism without the set encoder. The learning curves on four different tasks are plotted in Fig.~\ref{fig:set_comparison}. We find that the learning performance degrades when disabling the set encoder, indicating that the context-aware rank scores are beneficial to promote the performance. To qualitatively study the latent set embeddings from the set encoder, we randomly select some \emph{top-10} and \emph{last-10} ads from the test data set and generate their corresponding set embeddings $h'_i$ using the trained set encoder. The t-SNE~\cite{van2008visualizing} plots can be seen in Fig.~\ref{fig:tsne_plot}, where each point represents an ad. It is interesting to find that the top-ranked ads are clustered together, and the ``weak'' ads are separated from the cluster. This indicates that the set embedding $h'_i$ may carry the ``competitiveness'' information from the other candidate ads, assisting the subsequent rank score module to learn rank scores towards optimizing the overall performance.

% \footnote{By using t-SNE, the dimension of the set embedding $h'_i$ is reduced from $16$ to $2$.}

\section{Deployment in E-commerce Advertising System}
\label{app:deploy}
%   Then, the suitable candidate ads for the current PV request are selected and exposed to the user. Once the user finished interacting with the ad, either click, or place an order, these behaviors are recorded and sent to the log data collectors. A data pipeline behind these collectors will aggregate the event history into more valuable real-time processing and business log.
The Deep Neural Auction mechanism is deployed under ``\underline{A}dverti-sement \underline{I}ntelligent \underline{D}ecision-m\underline{A}king system'' (AIDA) in Taobao display advertising system. The online inference procedure can be formulated as follows:

 \begin{algorithm}
 \caption{Online Auction Service of DNA} \label{alg:alg_online}
 \begin{algorithmic}[1]
 \REQUIRE Online auction data (all candidate ads in a PV request), the trained DNA 
%  with temperature $\tau$ set to $0$
%  \STATE Extract feature set of candidate ads (such as pCTR, bid, etc)
 \STATE Data preprocessing: feature construction
 \STATE Generate rank scores of DNA with Eq.~(\ref{eq:deepset1})(\ref{eq:deepset2})(\ref{eq:minmax_forward})(\ref{eq:minmax_inv})
 \STATE Obtain the top-$K$ winning ads and their corresponding payments with differentiable sorting engine by setting $\tau=0$
 \end{algorithmic}
 \end{algorithm}

\begin{figure}[!ht]
\centering
\includegraphics[width=\linewidth]{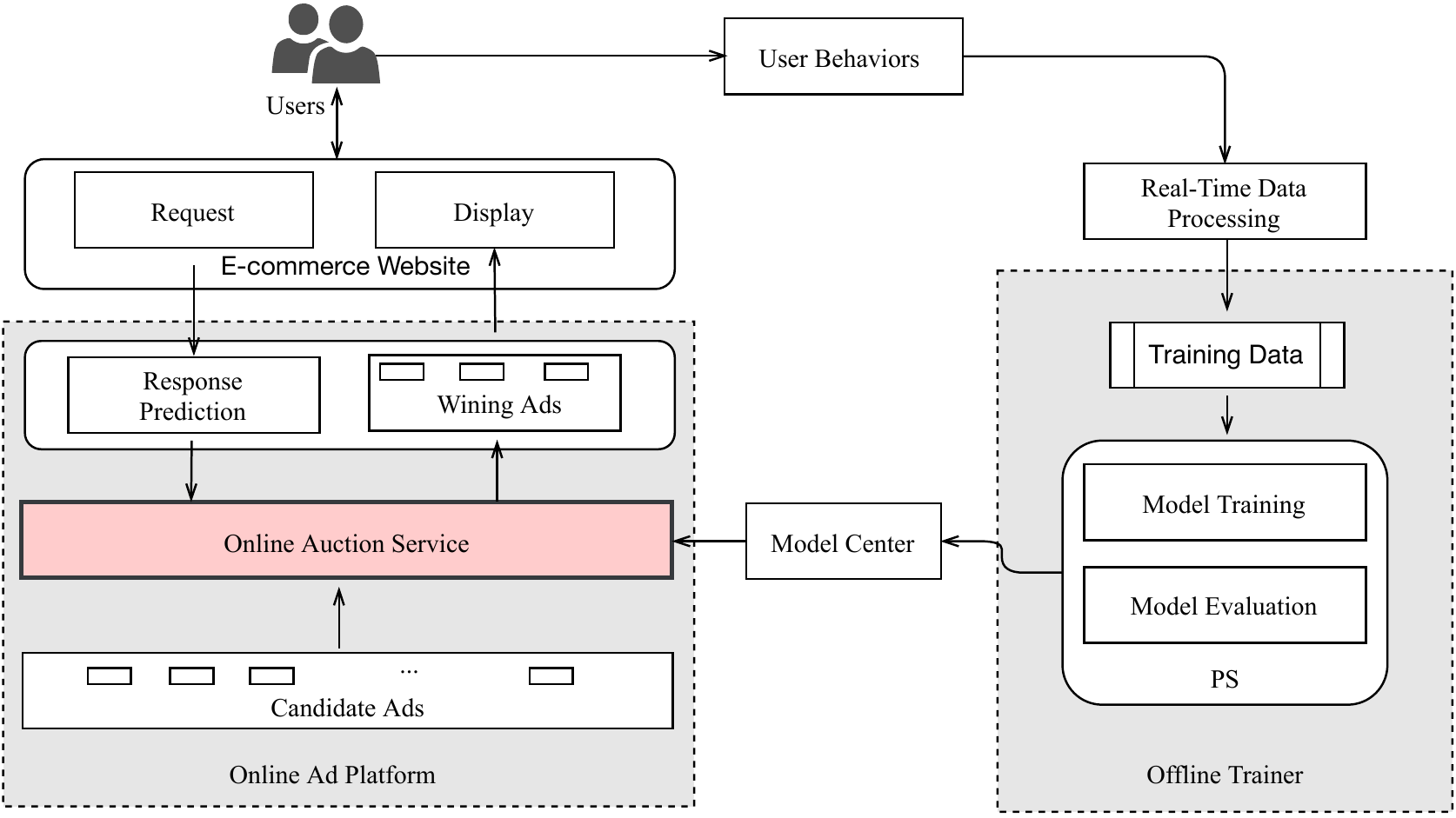}
\caption{The Pipeline of Training System.}
\label{fig:deployment-pipline}
\end{figure}

Fig.~\ref{fig:deployment-pipline} shows the overall architectures of the training system pipeline, including online ad platform and offline trainer. Firstly, the online ad platform receives a page view (PV) request from a user. The relevant candidate ads are selected, together with the generated user response predictions (\eg, $pCTR$, $pCVR$). Then, the auction mechanism is conducted and the top-$K$ winning ads will be selected and displayed to the user. Once the user finishes interacting with the ad, \eg, click, or place an order, these behaviors are recorded as log data and sent to the real-time data processing module, where hundreds of thousands of log data are processed per second. After that, the training procedure is performed via parameter-server (PS) framework and the model evaluation will be periodically carried on to monitor the training process. When the convergence condition is satisfied, the model checkpoint will be pushed to the model center module. The model checkpoint can be delivered in several minutes from offline to online. We also design a model version management tool inside the model center, endowing the training system with rollback ability in response to training crash or service breakdown. The whole pipeline from real-time data processing to model checkpoint transmission can be completed in less than 20 minutes.

% a request from the user is sent to the ad platform, where the relevant candidate ads are recalled and the corresponding user response prediction (\eg, pCTR, pCVR, etc.) are conducted. The top-$K$ winning ads will be selected in the online auction service based on the trained deep neural auction model. After seeing these ads, the user will interact with these ads, such as click the ads, place an order for the product, etc. These behaviors would be collected as log data. The real-time data processing module processes hundreds of thousands of log data per second and formats them as formatted samples for training. The model training module calculates the gradient of deep neural auction network parameters based on the parameter server (PS) framework. The trained model file will be immediately pushed to the model center module, which guarantees the model transmission from offline to online in millisecond delay and model version management with rollback capability. The pipeline from real-time feedback data to model training and online transmission will be completed in less than 20 minutes.

\begin{figure}[!ht]
\centering
\includegraphics[width=0.99\linewidth]{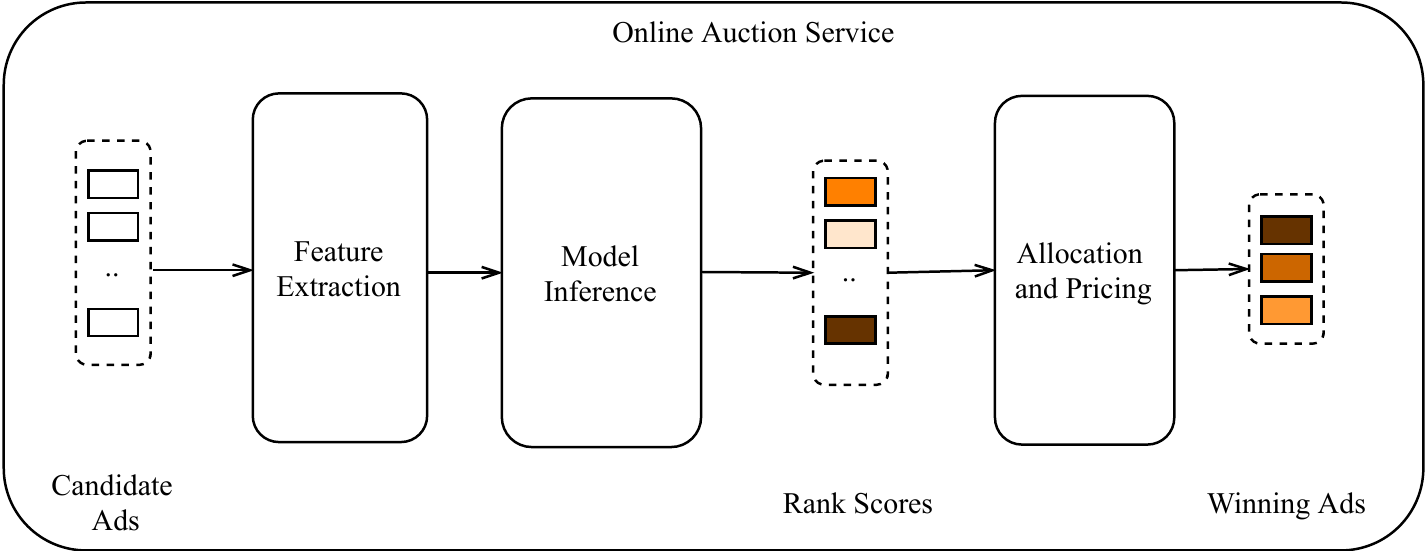}
\caption{Online Auction Service.}
% \vspace{-0.5em}m
\label{fig:deployment-online-service-module}
\end{figure}

The online auction service, as illustrated in Fig.~\ref{fig:deployment-online-service-module}, consists of three main components: a feature extraction module, a model inference module, and an allocation \& pricing module. Given the candidate ads set with the available information, the feature extraction module conducts feature engineering, which extracts useful information from the raw user and ads. The trained model checkpoint is then loaded and executed to generate rank scores for all ads in the model inference module. Finally, the allocation and payments are determined using the differentiable sorting engine in DNA (by setting $\tau=0$). We mainly focus on response time (RT) for the online auction service. The online auction service processes tens of thousands of ad auctions per second, and the RT is 6ms average with dozens of 32-CPU-cores servers. In the last Double Eleven shopping festival of \emph{Taobao}, the online auction service accommodated a tremendous nearly 1 million ad auctions per second during the peak, keeping all services in working order.

% and query-per-second (QPS)

% \section{Supplementary of Online A/B Testing}
% \label{app:online}

% \begin{figure}[tp]
% \centering
% \includegraphics[width=\linewidth]{pic/Online_ab_graph.pdf}
% \caption{Online A/B test compared with GSP on promoting (from Jan 26, 2021 to Feb 6, 2021, 1\% production flow.}
% \label{fig:Online_ab_graph}
% \end{figure}

% \begin{table}[!t]
%     \caption{Online A/B test compared with GSP on promoting (from Jan 26, 2021 to Feb 6, 2021, 1\% production flow).}
%     %\vspace{-0.6em}
%     \centering
%     \begin{tabular}{|c|c|c|c|c|c|c|}
%         \hline
%         \% Improved & Day 1 & Day 2 & Day 3 & Day 4 & Day 5 & Day 6 & Day 7 & Day 8 & Day 9 & Day 10 & Day 11 & Day 12\\
%         \hline
%         RPM & 7.23\% & 6.64\% & 5.51\%& 6.27\%& 6.00\%& 5.93\% & 7.05\% & 3.43\% & 5.82\% & 4.32\% & 6.33\% & 5.32\% \\
%         \hline
%         CTR & 21.18\% & 20.95\% & 16.87\% & 18.95\% & 17.46\% & 18.88\% & 18.83\% & 16.88\% & 18.45\% & 18.88\% & 19.79\% & 19.57\% \\
%         \hline
%         CVR & 23.63\% & 3.98\% & 22.94\% & 16.55\% & 21.90\% & 6.92\% & 11.96\% & 12.02\% & 10.27\% & 25.29\% & 6.83\% & 20.93\% \\
%         \hline
%         GPM & -3.79\% & 2.08\% & 10.77\% & 13.21\% & 16.59\% & 13.07\% & 32.64\% & 19.27\% & 10.31\% & 30.77\% & 23.94\% & 39.05\%\\
%         \hline
%     \end{tabular}
%     \label{tab:onlines_delete}
%     %\vspace{-0.6em}
% \end{table}

\end{document}